\documentclass[twocolumn]{aastex631}
\usepackage{amsmath}
\usepackage{chemformula}

\begin{document}

\title{Interstellar detection of O-protonated carbonyl sulfide, \ch{HOCS+}}


\author[0000-0001-9629-0257]{Miguel Sanz-Novo}
\affiliation{Centro de Astrobiolog{\'i}a (CAB), INTA-CSIC, Carretera de Ajalvir km 4, Torrej{\'o}n de Ardoz, 28850 Madrid, Spain}
\affiliation{Computational Chemistry Group, Departamento de Química Física y Química Inorgánica, Universidad de Valladolid, E-47011 Valladolid, Spain}

\author[0000-0002-2887-5859]{V\'ictor M. Rivilla}
\affiliation{Centro de Astrobiolog{\'i}a (CAB), INTA-CSIC, Carretera de Ajalvir km 4, Torrej{\'o}n de Ardoz, 28850 Madrid, Spain}

\author[0000-0003-4493-8714]{Izaskun Jim\'enez-Serra}
\affiliation{Centro de Astrobiolog{\'i}a (CAB), INTA-CSIC, Carretera de Ajalvir km 4, Torrej{\'o}n de Ardoz, 28850 Madrid, Spain}

\author[0000-0003-4561-3508]{Jes\'us Mart\'in-Pintado}
\affiliation{Centro de Astrobiolog{\'i}a (CAB), INTA-CSIC, Carretera de Ajalvir km 4, Torrej{\'o}n de Ardoz, 28850 Madrid, Spain}

\author[0000-0001-8064-6394]{Laura Colzi}
\affiliation{Centro de Astrobiolog{\'i}a (CAB), INTA-CSIC, Carretera de Ajalvir km 4, Torrej{\'o}n de Ardoz, 28850 Madrid, Spain}

\author[0000-0003-3721-374X]{Shaoshan Zeng}
\affiliation{Star and Planet Formation Laboratory, Cluster for Pioneering Research, RIKEN, 2-1 Hirosawa, Wako, Saitama, 351-0198, Japan}

\author[0000-0002-6389-7172]{Andr\'es Meg\'ias}
\affiliation{Centro de Astrobiolog{\'i}a (CAB), INTA-CSIC, Carretera de Ajalvir km 4, Torrej{\'o}n de Ardoz, 28850 Madrid, Spain}

\author[0000-0001-6049-9366]{\'Alvaro L\'opez-Gallifa}
\affiliation{Centro de Astrobiolog{\'i}a (CAB), INTA-CSIC, Carretera de Ajalvir km 4, Torrej{\'o}n de Ardoz, 28850 Madrid, Spain}

\author[0000-0001-5191-2075]{Antonio Mart\'inez-Henares}
\affiliation{Centro de Astrobiolog{\'i}a (CAB), INTA-CSIC, Carretera de Ajalvir km 4, Torrej{\'o}n de Ardoz, 28850 Madrid, Spain}

\author[0000-0002-7387-9787]{Sarah Massalkhi}
\affiliation{Centro de Astrobiolog{\'i}a (CAB), INTA-CSIC, Carretera de Ajalvir km 4, Torrej{\'o}n de Ardoz, 28850 Madrid, Spain}

\author[0000-0002-4782-5259]{Bel\'en Tercero}
\affiliation{Observatorio Astron\'omico Nacional (OAN-IGN), Calle Alfonso XII, 3, 28014 Madrid, Spain}

\author[0000-0002-5902-5005]{Pablo de Vicente}
\affiliation{Observatorio de Yebes (OY-IGN), Cerro de la Palera SN, Yebes, Guadalajara, Spain}

\author[0000-0001-7535-4397]{David San Andr\'es}
\affiliation{Centro de Astrobiolog{\'i}a (CAB), INTA-CSIC, Carretera de Ajalvir km 4, Torrej{\'o}n de Ardoz, 28850 Madrid, Spain}

\author[0000-0001-9281-2919]{Sergio Mart\'in}
\affiliation{European Southern Observatory, Alonso de C\'ordova 3107, Vitacura 763 0355, Santiago, Chile}
\affiliation{Joint ALMA Observatory, Alonso de C\'ordova 3107, Vitacura 763 0355, Santiago, Chile}

\author[0009-0009-5346-7329]{Miguel A. Requena-Torres}
\affiliation{University of Maryland, College Park, ND 20742-2421 (USA)}
\affiliation{Department of Physics, Astronomy and Geosciences, Towson University, Towson, MD 21252, USA}

\begin{abstract}

We present the first detection in space of O-protonated carbonyl sulfide (\ch{HOCS+}), in the midst of an ultradeep molecular line survey toward the G+0.693-0.027 molecular cloud. From the observation of all $K$$_a$ = 0 transitions ranging from $J$$_{lo}$ = 2 to $J$$_{lo}$ = 13 of \ch{HOCS+} covered by our survey, we derive a column density of $N$ = (9 $\pm$ 2)$\times$10$^{12}$ cm$^{-2}$, translating into a fractional abundance relative to H$_2$ of $\sim$7$\times$10$^{-11}$. Conversely, the S-protonated \ch{HSCO+} isomer remains undetected, and we derive an upper limit to its abundance with respect to H$_2$ of $\leq$3$\times$10$^{-11}$, a factor of $\geq$2.3 less abundant than \ch{HOCS+}. We obtain a \ch{HOCS+}/OCS ratio of $\sim$2.5$\times$10$^{-3}$, in good agreement with the prediction of astrochemical models. These models show that one of the main chemical routes to the interstellar formation of \ch{HOCS+} is likely the protonation of OCS, which appears to be more efficient at the oxygen end. Also, we find that high values of cosmic-ray ionisation rates (10$^{-15}$-10$^{-14}$ s$^{-1}$) are needed to reproduce the observed abundance of \ch{HOCS+}. In addition, we compare the O/S ratio across different interstellar environments. G+0.693-0.027 appears as the source with the lowest O/S ratio. We find a \ch{HOCO+}/\ch{HOCS+} ratio of $\sim$31, in accordance with other O/S molecular pairs detected toward this region and also close to the O/S solar value ($\sim$37). This fact indicates that S is not significantly depleted within this cloud due to the action of large-scale shocks, unlike in other sources where S-bearing species remain trapped on icy dust grains.

\end{abstract}
\keywords{Interstellar molecules(849), Interstellar clouds(834), Galactic center(565), Spectral line identification(2073), Astrochemistry(75)}

\section{Introduction} 
\label{sec:intro}

The pursuit of understanding the chemical reservoir of the interstellar medium (ISM) is a cornerstone of modern astrochemistry. At the heart of this quest is the investigation of molecules harboring sulfur (S), which play a key role in diverse aspects, from stellar nucleosynthesis and chemical evolution of galaxies \citep{perdigon21} to atmospheric chemistry in planets \citep{Krasnopolsky12,GomezMartin17,Chang23}, or biological processes \citep{Richard1986}, and they are considered essential ingredients for life on Earth \citep{Richardson13,Todd22}.

Starting with the discovery of carbon monosulfide (CS; \citealt{Penzias:1971kw}), the first S-bearing molecule observed in space, and followed by the detection of carbonyl sulfide (OCS; \citealt{Jefferts:1971hy}), more than thirty interstellar S-bearing molecules have been identified to date, which corresponds to $\sim$10 \% of the overall chemical inventory found in the ISM (see \citealt{McGuire22census} for a recent molecular census). In this context, recent detections of new S-bearing molecules (e.g., \ch{HC2S+}, \citealt{Cabezas22}; \ch{HC2S}, \ch{H2C2S}, \ch{H2C3S}, and \ch{C4S}, \citealt{Cernicharo21a}; \ch{HC3S+}, \citealt{Cernicharo21b}; \ch{CH3CH2SH}, \ch{HC(O)SH}, \citealt{Kolesnikova:2014fb,rodriguez-almeida2021a}; and HSO, \citealt{Marcelino23}), strongly encourage the astronomical community to hunt for new sulfurated candidates. However, in the case of S-containing species, additional factors need to be considered: i) despite being one of the most abundant elements in the Universe, its abundance is low compared to that of C and O ([S/O]$\sim$2.7$\times$10$^{-2}$, [S/C]$\sim$4.9$\times$10$^{-2}$; \citealt{Asplund:2009eu}); ii) S has a tendency to deplete rapidly onto the surface of interstellar dust grains \citep{JimenezEscobar:2014kt,Vidal17,Vidal18,Laas2019,shingledecker2020}, which dramatically decreases the abundances of S-bearing molecules in the gas phase \citep{Vidal17,Marcelino23,Fuente23}.

In recent years, searches for diverse S-bearing molecules of increasing complexity, such as thioformamide (\ch{NH2CHS}; \citealt{motiyenko20}), thioacetamide (\ch{CH3C(S)NH2}; \citealt{Remijan21}), thioacetaldehyde (\ch{CH3CHS}; \citealt{Margules20}), and methyl thioformate (\ch{CH3SC(O)H}, \citealt{Jabri20}) have remained so far unfruitful. All the above aspects aligns with the fact that almost the entire census of detected interstellar S-bearing molecules are S-analogues of the most abundant O-bearing species. Thus, we explore two S-analogues of the abundant protonated carbon dioxide (\ch{HOCO+}; \citealt{Thaddeus1981,Turner:1999yb,sakai_detection_2008,vastel_abundance_2016,Fuente16,Majumdar2018}), namely \ch{HOCS+} and \ch{HSCO+}, as auspicious candidates for interstellar detection. 

Moreover, pioneering works on the detection of protonated ions such as \ch{HCO+} \citep{Buhl:1970rd}, \ch{N2H+} \citep{Turner:1974jk}, and \ch{HCS+} \citep{Thaddeus1981} already proved that ion-molecule chemistry is efficient and takes place in the ISM \citep{Yamamoto17,Tinacci21,Puzzarini22}. This further strengthen the interest on both \ch{HOCS+} and \ch{HSCO+}, which may be formed via ion-molecule reaction starting from the ubiquitous carbonyl sulfide (OCS; \citealt{Goldsmith81,Li15,Boogert:2015fx}). Regarding the O-protonated isomer, \ch{HOCS+}, it has already been searched for toward different interstellar sources without success \citep{Turner1990Sulfur,Tercero:2010ft}.

In this work, we report the first interstellar detection of \ch{HOCS+} toward the Galactic Center molecular cloud G+0.693-0.027. We also search for the lower-in-energy isomer \ch{HSCO+} and derive an upper limit to its column density. Their relative abundance is presented and, by using gas-grain astrochemical models, we discuss the possible formation pathways of \ch{HOCS+}, that favor the production of this molecule over its isomer \ch{HSCO+}. Finally, we explore a sample of well-known S- and O-containing species detected toward several astronomical environments to shed light on the O/S ratio across the ISM. 

\section{Observations} 
\label{sec:obs}

We searched for both \ch{HOCS+} and \ch{HSCO+} isomers toward the molecular cloud G+0.693-0.027 (hereafter G+0.693), located in the Central Molecular Zone (CMZ) of the Milky Way. This astronomical source has been established as one of the prime targets to unravel new interstellar complex organic molecules (or COMs, defined as carbon-based molecules comprised of 6 or more atoms, \citealt{Herbst2020}), based on the first detection of a deluge of C-, O- and N- and also S-bearing species (see, e.g., \citealt{rivilla2019b,rivilla2020b,rivilla2021a,rivilla2021b,rivilla2022a,rivilla2022b,Rivilla23,Bizzocchi2020,rodriguez-almeida2021a,rodriguez-almeida2021b,jimenez-serra2022,zeng2021,zeng2023,SanzNovo23,Fatima23}).

We employed an unbiased ultradeep spectral survey, at sub-mK noise levels, carried out with the Yebes 40$\,$m (Guadalajara, Spain) and the IRAM 30$\,$m (Granada, Spain) radiotelescopes. We used the position switching mode, centered at $\alpha$ = $\,$17$^{\rm h}$47$^{\rm m}$22$^{\rm s}$, $\delta$ = $\,-$28$^{\circ}$21$^{\prime}$27$^{\prime\prime}$, with the off position shifted by $\Delta\alpha$~=~$-885$$^{\prime\prime}$ and $\Delta\delta$~=~$290$$^{\prime\prime}$. We used Yebes 40$\,$m observations to cover the $Q$-band (31.075-50.424 GHz), and IRAM 30$\,$m observations to cover three frequency windows: 83.2$-$115.41 GHz, 132.28$-$140.39 and 142$-$173.81 GHz. More detailed information of the ultradeep spectral survey (e.g., resolution and noise levels of the spectra), including the new Yebes 40$\,$m (project 21A014; PI: Rivilla) and IRAM 30$\,$m observations (project 123-22; PI: Jim\'enez-Serra) can be found in \citet{Rivilla23} and \citet{SanzNovo23}. Note that the sensitivity of this molecular line survey has seen a significant improvement compared to previous works (e.g., \citealt{rivilla2021a,rodriguez-almeida2021a,jimenez-serra2022}), reaching now sub-mK noise levels.

\section{Analysis and results} 
\label{sec:res}

\subsection{Rotational spectroscopy considerations} 
\label{subsec:rotspec}

\begin{center}
\begin{figure*}[ht]
     \centerline{\resizebox{1.0
     \hsize}{!}{\includegraphics[angle=0]{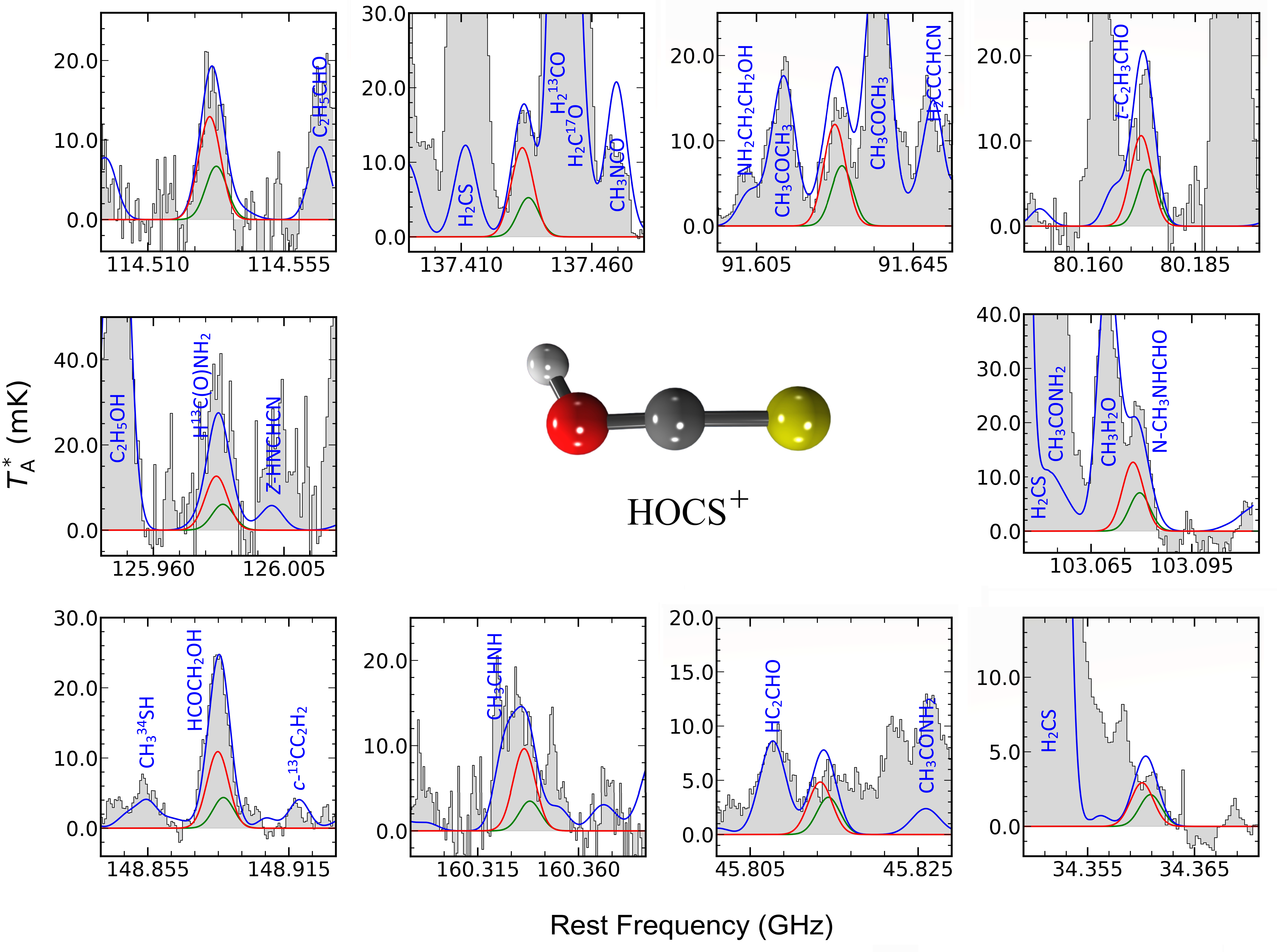}}}
     \caption{Transitions of \ch{HOCS+} identified toward the G+0.693–0.027 molecular cloud (listed in Table \ref{tab:oprot}). The result of the best LTE fit of \ch{HOCS+} is shown with a red line, the green line plots the predicted emission of HNC$^{34}$S, and the blue line plots the emission from all the molecules identified to date in our survey (including the latter two). The observed spectra are plotted as gray histograms. The structure of \ch{HOCS+}, taken from \citet{Fortenberry2012}, is also shown (carbon atoms in gray; oxygen atoms are in red, sulfur atoms in yellow and hydrogen atoms in white). Note that HNC$^{34}$S is blended with \ch{HOCS+} because of their similarity of $B$ + $C$ but its contribution can be well constrained based on the HNC$^{32}$S/HNC$^{34}$S isotopic ratio (see text).}
\label{f:LTEspectrum}
\end{figure*}
\end{center}

The [H,C,S,O]$^+$ isomeric family comprises two structural isomers\footnote{Molecular species with the same number of atoms of each element, but arranged with different bonds between them.} of remarkable stability, the S-protonated carbonyl sulfide isomer, \ch{HSCO+}, and the O-protonated form, \ch{HOCS+}, which was predicted to lie at 4.9 kcal mol$^{-1}$ (2466 K) higher in energy than \ch{HSCO+} \citep{Wheeler06}. Regarding their experimental spectroscopic characterization, these two isomeric arrangements were studied previously in the centimeter-wavelengths by \citet{Ohshima96} and \citet{McCarthy07}, who characterized \ch{HOCS+} and \ch{HSCO+}, respectively, using high-resolution rotational spectroscopy. An additional cyclic structure, \ch{OC(H)S+}, was also located in the potential energy surface (PES) lying at 56.05 kcal mol$^{-1}$ (30219 K) above \ch{HSCO+} (at the CCSD(T)/cc-pVQZ level of theory; \citealt{Wheeler06}), although it remains undetected in the laboratory.

\begin{table*}
\centering
\tabcolsep 3pt
\caption{Spectroscopic information of the selected transitions of \ch{HOCS+} detected toward G+0.693$-$0.027 (shown in Figure \ref{f:LTEspectrum}).}
\begin{tabular}{ccccccccccc}
\hline\hline
Frequency & Transition $^{(a)}$ & log \textit{I}& \textit{g}$\mathrm{_u}$ & $E$$\mathrm{_{up}}$ &  Blending  \\ 
(GHz) & &  (nm$^2$ MHz) &  &  (K)  & \\
\hline
34.3598460  & 3$_{0,3}$ -- 2$_{0,2}$  & --5.0106  & 7  & 3.3 &  \ch{HNC$^{34}$S} \\ 
45.8130061*  & 4$_{0,4}$ -- 3$_{0,3}$  & --4.6386  & 9  & 5.5   & \ch{HNC$^{34}$S} \\
80.1717639*  & 7$_{0,7}$ -- 6$_{0,6}$  & --3.9226  & 15  &  15.3 & \ch{HNC$^{34}$S} and \ch{\textit{t}-C2H3CHO} \\
91.6243553*  & 8$_{0,8}$ -- 7$_{0,7}$  & --3.7546  & 17  & 19.7 & \ch{HNC$^{34}$S} \\
103.0767395*  & 9$_{0,9}$ -- 8$_{0,8}$  & --3.6079  & 19 &  24.6 & \ch{HNC$^{34}$S} and \ch{N-CH3NHCHO}\\
114.5288908*  & 10$_{0,10}$ -- 9$_{0,9}$  & --3.4782  & 21  &  30.0 & \ch{HNC$^{34}$S} \\
125.9807832*  & 11$_{0,11}$ -- 10$_{0,10}$  & --3.3624  & 23 &  36.0  & \ch{HNC$^{34}$S} and \ch{H$^{13}$CONH2}  \\
137.4323908*  & 12$_{0,12}$ -- 11$_{0,11}$  & --3.2582  & 25  & 42.6 & \ch{HNC$^{34}$S} \\
148.8836878*  & 13$_{0,13}$ -- 12$_{0,12}$  & --3.1639  & 27  & 49.7 & \ch{HNC$^{34}$S} and \ch{HCOCH2OH} \\
160.3346483*  & 14$_{0,14}$ -- 13$_{0,13}$  & --3.0781  & 29  & 57.3 & \ch{HNC$^{34}$S} and \ch{\textit{Z}-CH3CHNH} \\
\hline 
\end{tabular}
\label{tab:oprot}
\vspace*{-2.5ex}
\tablecomments{$^{(a)}$ The rotational energy levels are labelled using the conventional notation for asymmetric tops: $J_{K_{a},K_{c}}$, where $J$ denotes the angular momentum quantum number, and the $K_{a}$ and $K_{c}$ labels are projections of $J$ along the $a$ and $c$ principal axes. Lines that have been observed for the first time in the present astronomical dataset are marked with a * symbol.}
\end{table*}

According to previous theoretical structural studies \citep{Wheeler06,Fortenberry2012} both isomers are asymmetric tops near the prolate limit (see e.g., Figure \ref{f:LTEspectrum} for \ch{HOCS+}). \ch{HOCS+} presents a slightly bent heavy atom skeleton, with $\angle$(O-C-S) = 174.4$^{\circ}$ \citep{Fortenberry2012}, while the H atom is located at $\angle$(H-O-C) = 117.9$^{\circ}$ in the $ab$ plane, which implies that \textit{b}-type lines are also allowed, in principle, by dipole moment selection rules, even though there is no spectroscopic evidence for them in previous experiments (due to the large value of the $A$ rotational constant, $A$ = 782696 MHz). Therefore, the $b$-type spectrum of \ch{HOCS+} remains unknown. \cite{Ohshima96} identified the three lowest-$J$ $R$-branch $a$-type rotational transitions belonging to the $K$$_a$ = 0 ladder of \ch{HOCS+}, which were re-measured later on by \citet{Gottlieb:2000nc}. Therefore, these single series of lines will be the main target of our astronomical search. Note that this progression can be easily extrapolated to higher frequencies, although larger uncertainties will be found once reaching the millimeter-wave region (e.g. 0.3 MHz at 100 GHz and 0.6 MHz at 130 GHz, which translates into 1.1 and 1.8 km s$^{-1}$, respectively). Nevertheless, even at these frequencies, the uncertainties in the extrapolation are considerably smaller than the standard line widths of the molecular line emission measured toward G+0.693 (FWHM $\sim$ 15$-$20 km s$^{-1}$; \citealt{requena-torres_organic_2006,requena-torres_largest_2008,zeng2018}). Consequently, they will not have an impact on the present analysis.

\subsection{Detection of \ch{HOCS+} and search for \ch{HSCO+}} 
\label{subsec:detection}

We used the Cologne Database for Molecular Spectroscopy (CDMS) entry 061510 \citep{Muller2005, endres2016} and performed the astronomical line identification of \ch{HOCS+} using the Spectral Line Identification and Modeling (SLIM) tool (version from 2023 November 15) within the \textsc{Madcuba} package \citep{martin2019}. This tool works assuming a Local Thermodynamic Equilibrium (LTE) excitation and enable us to generate the LTE synthetic spectra. In Figure \ref{f:LTEspectrum} we depict all the transitions of \ch{HOCS+} that fall within the current astronomical dataset, which are detected toward G+0.693. Their spectroscopic information is listed in Table \ref{tab:oprot}. The fitted line profiles of \ch{HOCS+} are depicted with a red solid line overlaid with the observed spectra (in gray). \ch{HOCS+} appears to be the dominant carrier of the spectral features, despite the observed partial blend with mainly transitions of HNC$^{34}$S, which will be thoroughly explained below. Also, we stress that we have identified the full progression of $K$$_a$ = 0 lines spanning from $J$$_{lo}$ = 2 (measured in the $Q$-band) to $J$$_{lo}$ = 13 (in the 2mm atmospheric window), with the exception of the $J$ = 4,5 lines which are not observable due to atmospheric opacity. Interestingly, apart from the 3$_{0,3}$ -- 2$_{0,2}$ transition, the rest of the lines have been observed for the first time in the spectral survey of G+0.693 and still remain undetected in the laboratory (see Table \ref{tab:oprot}).

In order to confirm that the spectral features we observe arise mainly from \ch{HOCS+} and to thoroughly assess potential line contamination from other molecules, we have taken into account the emission profiles of the more than 130 molecules previously identified toward G+0.693 (\citealt{Rivilla23} and references therein). We found that most of the lines of \ch{HOCS+} are partially blended with the emission from the $^{34}$S isotopologue of HNCS (HNC$^{34}$S), which to our best knowledge has not been detected yet in the ISM. By coincidence, the $B$ and $C$ rotational constants of both species are extremely similar ($B$+$C$ = 11453.3 MHz for \ch{HOCS+} versus 11453.6 MHz for HNC$^{34}$S; \citealt{Yamada:1980ky}) and, therefore, their $a$-type spectra will be governed by nearly identical patterns. To accurately model their expected emission, we can benefit from two different facts: i) the $^{32}$S/$^{34}$S isotopic ratio for HNC$^{32}$S/HNC$^{34}$S, which will provide information on the expected molecular column density ($N$) of HNC$^{34}$S; ii) the different intensities of the rotational lines of both molecules, which will also appear as a fingerprint for each species.

Following this approach, to estimate the contribution of HNC$^{34}$S we have used the $^{32}$S/$^{34}$S isotopic ratio derived for a variety of molecules (CS, OCS, CCS, SO and \ch{H2S}) and using multiple isotopologues (Colzi et al. in prep.). These values are always $\geq$20 toward G+0.693, in good accordance with the Solar System $^{32}$S/$^{34}$S ratio of $\sim$22 derived by \citet{wilson_isotopes_1999}. They are also close to the value obtained toward the envelope of the neighbouring region Sgr B2(N) by \cite{Humire2020} ($\sim$18), and that reported recently by \cite{Li2023} (17 $\pm$ 1). We then choose the lowest $^{32}$S/$^{34}$S ratio ($\sim$20), which corresponds to the ratio obtained for the possible precursor of \ch{HOCS+}, OCS (see Sect. \ref{subsec:detectionOCS} and Appendix \ref{AnalysisOCS}), and also to that derived for CS (20 $\pm$ 2; Colzi et al in prep.). This is the most conservative or worst case scenario, which is the one employed to produce Figure \ref{f:LTEspectrum}. Larger values (i.e. 30-40) will imply a lower column density for HNC$^{34}$S and, therefore, a lower contribution of HNC$^{34}$S to the observed profiles and more ``available" emission to be fit by \ch{HOCS+}, which will be less contaminated (see below). After considering the column density of the parent species, HNCS (analyzed in the Appendix \ref{AnalysisHNCS}), we produced the LTE synthetic spectra of HNC$^{34}$S adopting the same physical parameters as derived for the $^{32}$S isotopologue (i.e., excitation temperature of $T_{\rm ex}$ = 20.4 K, radial velocity of $v$$_{\rm LSR}$ = 66.7 km s$^{-1}$ and line width of FWHM = 21.0 km s$^{-1}$) and a $N$(HNC$^{34}$S) = 3.1 $\times$10$^{12}$ cm$^{-2}$. The possibility that HNC$^{34}$S alone is the molecular carrier of the observed lines was rapidly ruled out, since a significantly larger $N$(HNC$^{34}$S), which implies a HNC$^{32}$S/HNC$^{34}$S ratio $\sim$7, almost a factor of $\sim$3 lower than the value derived for other molecules, was needed to properly explain all the observed emission.

\begin{table*}
\centering
\caption{Derived physical parameters for OCS, OC$^{34}$S, HNCS, HNC$^{34}$S, \ch{HOCO+}, \ch{HOCS+} and \ch{HSCO+} toward the G+0.693-0.027 molecular cloud.}
\begin{tabular}{ c c c c c c c  }
\hline
\hline
 Molecule & Formula & $N$   &  $T_{\rm ex}$ & $v$$_{\rm LSR}$ & FWHM  & Abundance$^a$   \\
 & & ($\times$10$^{14}$ cm$^{-2}$) & (K) & (km s$^{-1}$) & (km s$^{-1}$) & ($\times$10$^{-10}$)    \\
\hline
Carbonyl sulfide & OCS & 36.1 $\pm$ 0.5 & 22.9 $\pm$ 0.3 & 66.8 $\pm$ 0.1 & 21.8 $\pm$ 0.3 & 269 $\pm$ 24  \\
Carbonyl sulfide ($^{34}$S isotopologue) & OC$^{34}$S & 1.80 $\pm$ 0.05 & 23.5 $\pm$ 0.6 & 66.7 $\pm$ 0.3 & 23.9 $\pm$ 0.7 &  13.4 $\pm$ 1.2   \\
\hline
Isothiocyanic acid  & HNCS & 0.62 $\pm$ 0.01 & 20.4 $\pm$ 0.5 & 66.7 $\pm$ 0.3 & 21.0$^b$ & 4.63 $\pm$ 0.41  \\
Isothiocyanic acid ($^{34}$S isotopologue)  & HNC$^{34}$S & $\sim$0.031$^c$  & 20.4$^b$ & 66.7$^b$ & 21.0$^b$ & 0.23$^c$   \\
\hline
O-protonated carbonyl sulfide & \ch{HOCS+} & 0.09 $\pm$ 0.02 & 28$^b$  & 66.8$^b$ & 21.8$^b$ &  0.7 $\pm$ 0.2   \\
S-protonated carbonyl sulfide & \ch{HSCO+} & $\leq$ 0.04  & 28$^b$ & 66.8$^b$ & 21.8$^b$ & $\leq$ 0.3  \\
\hline 
Protonated carbon dioxide ($K$$_a$ = 0) & \ch{HOCO+}  &  2.83 $\pm$ 0.03  & 11.4 $\pm$ 0.1  & 66.9 $\pm$ 0.1  & 21.0  $\pm$ 0.2 & 21.1  $\pm$ 1.9   \\
\hline
\end{tabular}
\label{tab:comparison}
\vspace{0mm}
\vspace*{-1.5ex}
\tablecomments{$^a$ We adopted $N_{\rm H_2}$ = 1.35$\times$10$^{23}$ cm$^{-2}$, from \citet{martin_tracing_2008}, assuming an uncertainty of 15\% of its value. $^b$ Value fixed in the fit. $^c$ Derived by applying the $^{32}$S/$^{34}$S=20 ratio to $N$(HNCS).}
\label{tab:g0693}
\end{table*}

\begin{figure}
\centerline{\resizebox{0.95\hsize}{!}{\includegraphics[angle=0]{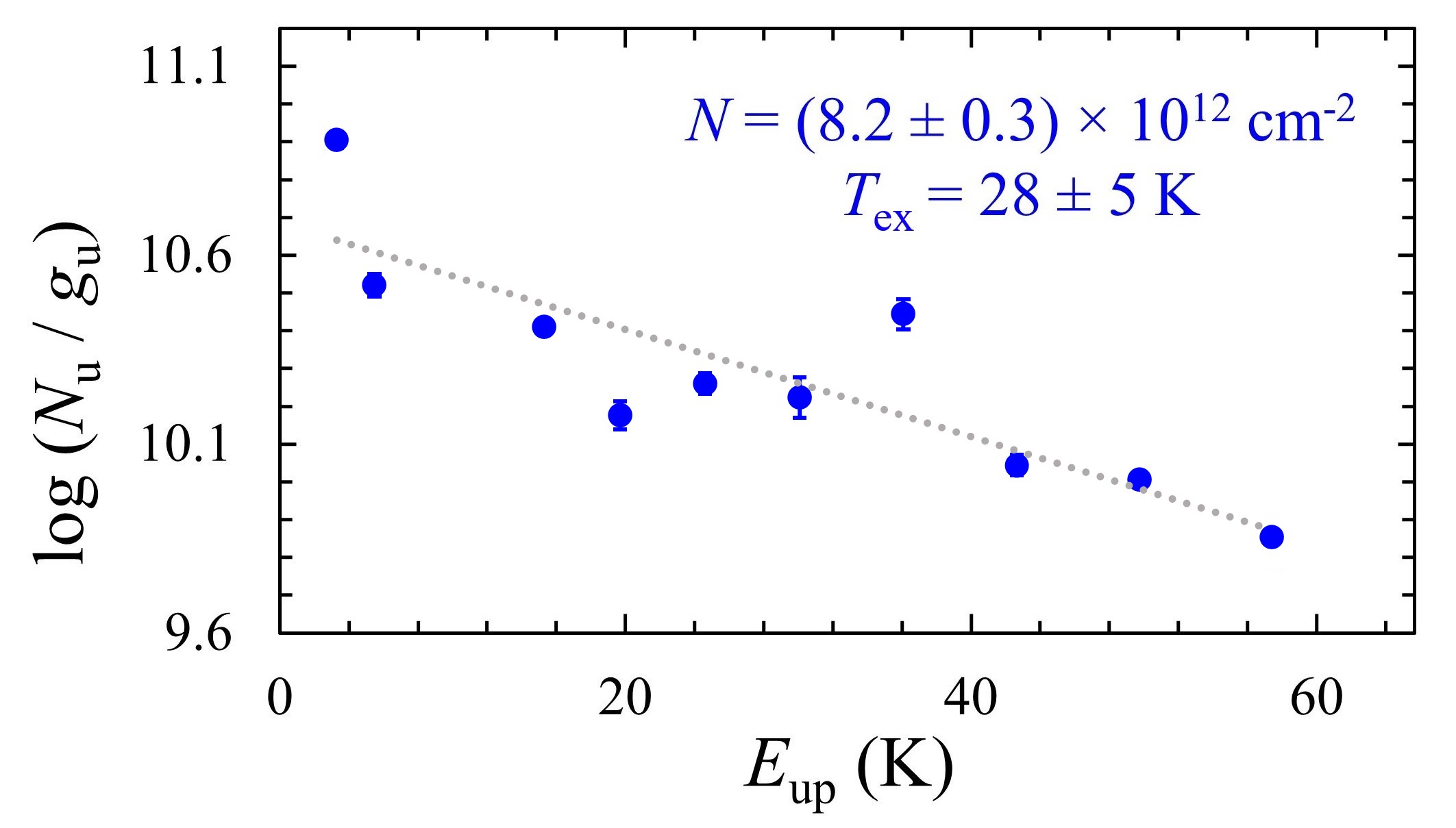}}}
\caption{Population diagram of \ch{HOCS+} toward G+0.693 (depicted in blue dots). The gray dotted line corresponds to the best linear fit to the data points. The derived values for the molecular column density, $N$ and the excitation temperature, $T_{\rm ex}$, are shown in blue. Note that the contamination by all other identified species has been removed from the observed data.} 
\label{f:rotdiagram}
\end{figure}

Regarding the LTE analysis of \ch{HOCS+}, we initially used the same parameters of $T_{\rm ex}$ = 22.9 K, $v$$_{\rm LSR}$ = 66.8 km s$^{-1}$ and FWHM = 21.8 km s$^{-1}$ derived from its proposed precursor, OCS, listed in Table \ref{tab:comparison} (see Sect. \ref{subsec:detectionOCS} and Appendix \ref{AnalysisOCS}). {We then performed a population or rotational diagram analysis \citep{goldsmith1999}, as implemented in \textsc{Madcuba}, using all the transitions listed in Table \ref{tab:oprot} and the velocity-integrated intensity over the line width \citep{rivilla2021a}. The latest version of SLIM also offers the possibility of considering the emission from other identified species to further remove it from the observed data. The results are shown in Figure \ref{f:rotdiagram}. Following this approach, we derived the following physical parameters for \ch{HOCS+}: $N$ = (8.2 $\pm$ 0.3) $\times$10$^{12}$ cm$^{-2}$ and $T_{\rm ex}$ = 28 $\pm$ 5 K. Afterward, we carried out the LTE fit to the \ch{HOCS+} emission using the} \textsc{Autofit} tool within SLIM \citep{martin2019}, which performs a nonlinear least-squares LTE fit to the observed spectra, with the column density left as a free parameter and fixing the $T_{\rm ex}$ to the value obtained with the rotational diagram (28 K). We employed again all the transitions listed in Table \ref{tab:oprot} and accounted for the expected emission from every molecule detected within the same frequency interval. The results of the best LTE fit are presented in Table \ref{tab:comparison}. We derived a molecular column density of $N$ = (9 $\pm$ 2) $\times$10$^{12}$ cm$^{-2}$, which yields a fractional abundance with respect to molecular hydrogen of (7 $\pm$ 2) $\times$ 10$^{-11}$, using a $N$(H$_{2}$) = 1.35$\times$10$^{23}$ cm$^{-2}$ from \citet{martin_tracing_2008}. Moreover, we stress that if we consider a larger $^{32}$S/$^{34}$S ratio, such as 40 instead of 20 to estimate the contribution of HNC$^{34}$S, the fit would yield a $N$(\ch{HOCS+}) = (1.1 $\pm$ 2) $\times$10$^{13}$ cm$^{-2}$, just a factor of $\sim$1.2 higher. Alternatively, if we adopt the $T_{\rm ex}$ of OCS (23 K) to perform the \textsc{Autofit}, it only yields a change in the molecular abundance of $\sim$10 \%, which is within the errors of the derived $N$, but slightly overestimates the emission of the 3$_{0,3}$ -- 2$_{0,2}$ and 4$_{0,4}$ -- 3$_{0,3}$ transitions.

\begin{figure*}
\centerline{\resizebox{0.95\hsize}{!}{\includegraphics[angle=0]{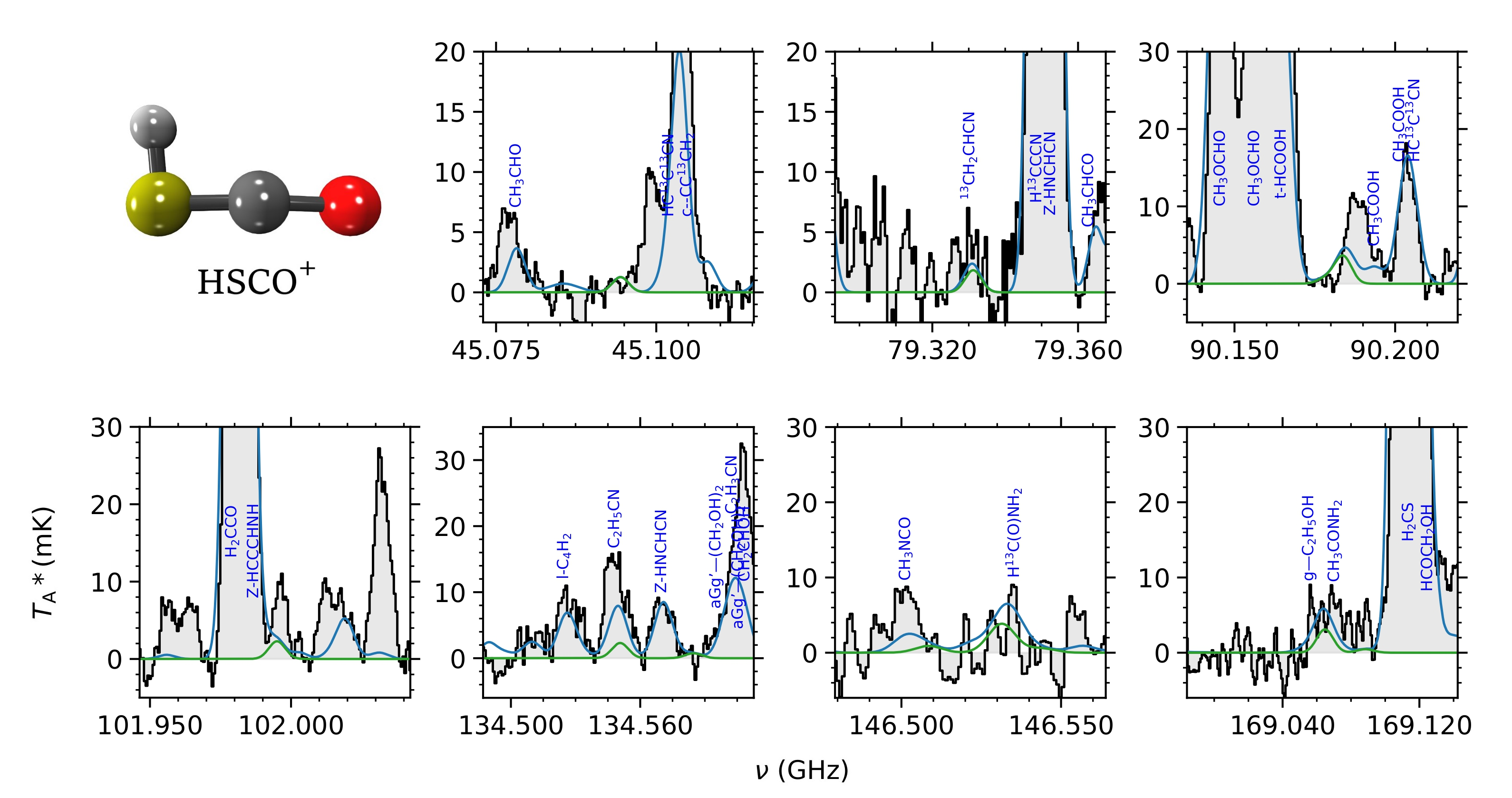}}}
\caption{LTE simulation of the \ch{HSCO+} emission at the 3$\sigma$ upper limit column density derived toward G+0.693 using the physical parameters shown in Table \ref{tab:comparison} (in green) together with the expected molecular emission from all the molecular species identified to date in our survey (in blue), both overlaid on the observations (black line and in gray histogram). The structure of \ch{HSCO+} is also shown.}
\label{f:LTEspectrumlimit}
\end{figure*}

During the analysis, we have also tried to include the newly observed astronomical lines of \ch{HOCS+} in a global fit to a semirigid rotor Hamiltonian along with the previously measured laboratory lines. However, the typical broad line widths of G+0.693, together with the observed partial blends detailed above, did not enable us to determine further centrifugal distortion parameters and improve the set of spectroscopic constants of the molecule (as was done successfully for carbonic acid, HOCOOH; \citealt{SanzNovo23}). We also inspected the observational data searching for new $K$$_a$ = 1 transitions, but these attempts were unfruitful due to the lack of reliable laboratory measurements for higher $K$$_a$ ladders. Thus, we encourage the laboratory spectroscopic community to perform additional high-resolution rotational measurements and extend the frequency coverage for \ch{HOCS+} up to the millimeter and submillimeter-wave region. In particular, the detection of higher $K$$_a$ ladders will aid to constrain more accurately both $B$ and $C$ rotational constants as well as centrifugal distortion parameters, thereby enabling us to further disentangle the emission of \ch{HOCS+} and HNC$^{34}$S.

Finally, we searched for the lower-in-energy S-protonated isomer, \ch{HSCO+}, using the high-resolution rotational data reported in the multi-technique spectroscopic study of \citet{Lattanzi18} together with the previous laboratory data from \citet{McCarthy07}, which corresponds to the 061509 entry of the CDMS catalog \citep{Muller2005}. In this case, extensive millimeter and submillimeter data were reported by \citet{Lattanzi18}, which allow us to extend the reliability of the predictions for the S-protonated \ch{HSCO+} to transitions with higher $K$$_a$ values. This isomer remains undetected in the present observational data (only two almost unblended transitions are found, as shown in Figure \ref{f:LTEspectrumlimit}), so we derived an upper limit to its column density. We employed the same physical parameters obtained for the O-protonated form (see Table \ref{tab:comparison}) and selected the brightest $a$-type $R$-branch $K$$_a$ = 0,1 transition that appears clean from the emission from other molecules, which correspond to the 3$_{0,3}$ -- 4$_{0,4}$ (located at 45.094116 GHz) rotational transition. We derived a line-integrated 3$\sigma$ upper limit to its column density of $N$ $\leq$ 4$\times$10$^{12}$ cm$^{-2}$, which translates into an upper limit to the molecular abundance relative to H$_{2}$ of $\leq$3$\times$10$^{-11}$. Consequently, it is at least a factor of $\geq$2.3 less abundant than \ch{HOCS+}.

\subsection{Analysis of OCS and OC$^{34}$S} 
\label{subsec:detectionOCS}

For completeness, we also modeled the emission of the related carbonyl sulfide (OCS, $v$ = 0), which appears as a promising precursor of \ch{HOCS+}, and its $^{34}$S monosubstituted isotopologue OC$^{34}$S. We employed the entries 060503 and 062505, respectively, of the CDMS catalog. The details of the analysis are reported in the Appendix \ref{AnalysisOCS} (see Tables \ref{tab:ocs} and \ref{tab:ocs34}, and Figures \ref{f:OCS} and \ref{f:OCS34}). In Table \ref{tab:comparison}, we present the results for the best-fitted LTE model for both species using the \textsc{Autofit} tool within SLIM. We obtained a OC$^{32}$S/OC$^{34}$S ratio of 20 $\pm$ 1 and a \ch{HOCS+}/\ch{OCS} ratio of $\sim$2.5$\times$10$^{-3}$ toward G+0.693.

\subsection{Analysis of \ch{HOCO+}} 
\label{subsec:detectionHOCO}

In order to compare these results with the structurally similar O-bearing molecules, we also report the detection of the ground state of \ch{HOCO+}, the O-analogue of \ch{HOCS+}, toward G+0.693. To carry out its interstellar search, we used the rotational data reported in \citealt{Bizzocchi17} (entry 045522 of the CDMS catalog). The LTE analysis is tackled in detail in the Appendix \ref{AnalysisHOCO} (see Table \ref{tab:hoco} and Figure \ref{f:LTEspectrumHOCO}), while the results of the \textsc{Autofit} are presented in Table \ref{tab:comparison} along with the physical parameters of \ch{HOCS+} and \ch{HSCO+}. We derived a \ch{HOCO+}/\ch{HOCS+} ratio of 31 $\pm$ 6 toward this source, which will be discussed in Sect. \ref{subsec:soratio}

\section{Discussion} 
\label{sec:disc}

\subsection{\ch{HOCS+}/\ch{HSCO+} isomeric ratio}
\label{subsec:othermolecules}

The identification of the higher-energy O-protonated isomer, \ch{HOCS+}, and the nondetection of the S-protonated form, \ch{HSCO+} (global minimum in energy), toward G+0.693 raises an interesting question: Why is the most stable isomer not being detected? Experimentally, it is found that \ch{HOCS+} is less abundant than \ch{HSCO+}, obtaining a ratio of $\sim$1:3 in the microwave spectrum \citep{McCarthy07}. This result approximately correlates with the theoretically computed energy difference between both isomers \citep{Wheeler06}. Note that the population observed in the supersonic jet during the experiment \citep{McCarthy07} is related to that in an equilibrium situation prior to the adiabatic expansion (at relatively high temperatures of $\sim$300 K). However, our observational findings points to an opposite trend, showing a lower limit to the \ch{HOCS+}/\ch{HSCO+} ratio of $\geq$2.3. The relative population must then depend on kinetic factors rather than the relative thermodynamic stability of the S- and O-protonated isomers, which will be discussed in Sect. \ref{subsec:formation}. Hence, these results are an additional example against the minimum energy principle (MEP; \citealt{Lattelais:2011bo}), which proposes that the thermodynamically favored (most stable) isomer is expected to be predominant in the ISM. This fact is not striking, since the MEP has been proven to fail to predict the relative abundances of a plethora of structural isomers (e.g., the \ch{C3H2O}, \ch{C2H4O2} or \ch{C2H5O2N} isomeric families; \citealt{shingledecker2019,mininni2020,Rivilla23}). Thus, the identification of \ch{HOCS+} provides clear observational evidence to argue that the MEP does not apply to the interestellar [H,C,S,O]$^+$ isomeric family either. 


\subsection{Interstellar formation of \ch{HOCS+}}
\label{subsec:formation}

\begin{figure}
\centerline{\resizebox{1\hsize}{!}{\includegraphics[angle=0]{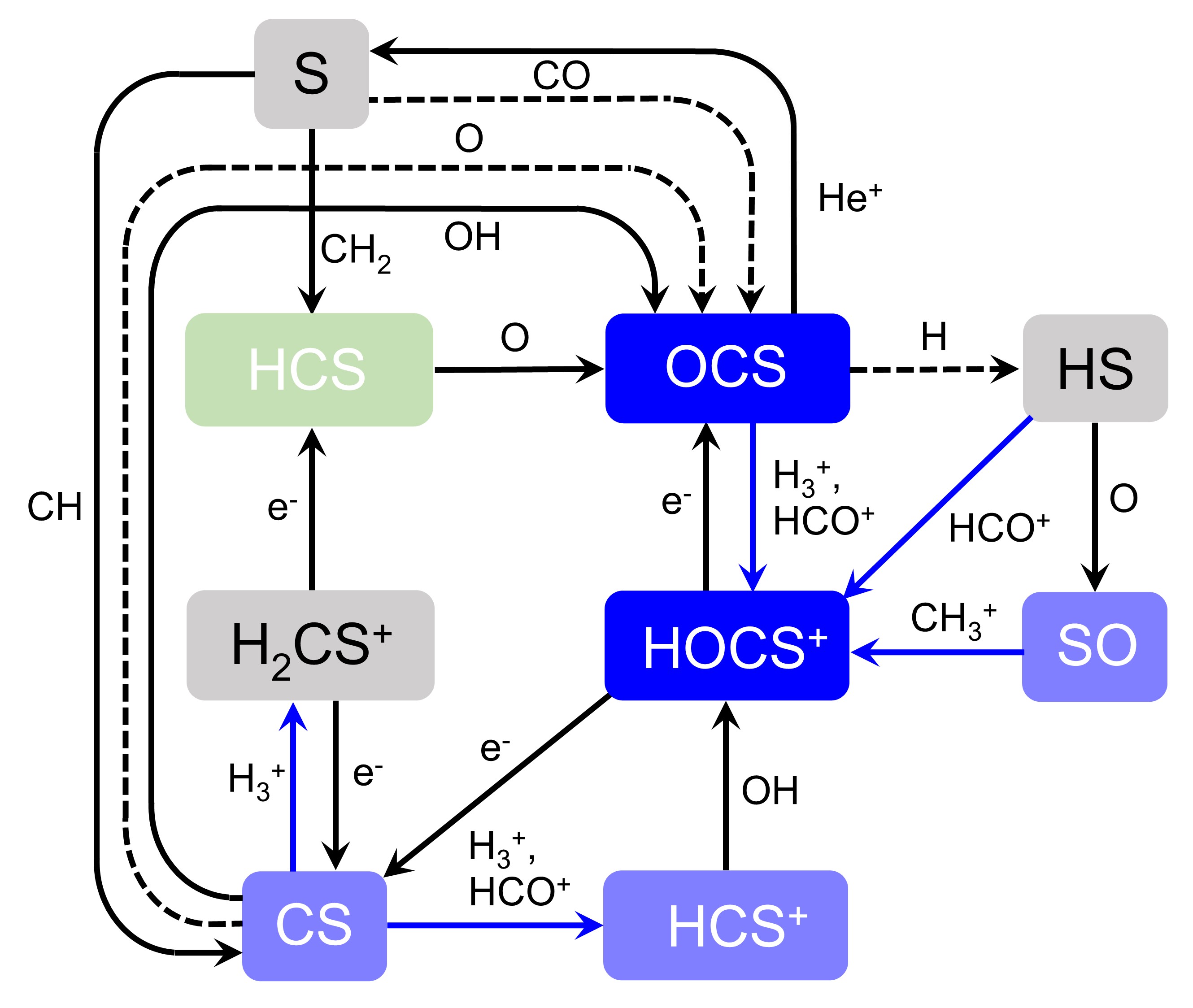}}}
\caption{Summary of the main chemical routes for the formation of \ch{HOCS+} in the ISM, adapted and completed from \citet{Turner1990Sulfur}. The main formation route (see discussion in the text) is highlighted with dark blue bloxes. We depict in blue the molecules that have been identified toward the G+0.693 molecular cloud, in green those species that have been searched for but not detected toward G+0.693, and in gray the molecules that have not been searched for toward G+0.693 because spectroscopy is not available (\ch{H2CS+}) or no transitions fall in the spectral survey (S and HS). Surface reactions are shown in black dashed arrows, gas-phase reactions are indicated in black solid lines and gas-phase protonation reactions are shown with blue arrows.}
\label{f:model}
\end{figure}

Several reactions have already been proposed to produce at least one protonated carbonyl sulfide isomer \citep{Turner1990Sulfur,Liu02,Vidal17,Tinacci21}. Owing to the ubiquitous nature of OCS and its relatively large abundance in the ISM, chemical intuition suggests that a direct ion-molecule reaction, using protons from diverse sources, may be the dominant formation route. For example, they can form through:

\begin{equation}
\mathrm{OCS} + \mathrm{\ch{H3+}} \rightarrow \rm \mathrm{\ch{HOCS+}} + \mathrm{\ch{H2}} 
\label{eq1}
\end{equation}
\begin{equation}
\mathrm{OCS} + \mathrm{\ch{HCO+}} \rightarrow \rm \mathrm{\ch{HOCS+}} + \mathrm{\ch{CO}} 
\label{eq2}
\end{equation}

which will be the most promising routes to form these isomers (the complete reaction scheme is shown in Figure \ref{f:model}). Regarding OCS, it can be formed efficiently on the grains via CO + S \citep{shingledecker2020}, being then released into the gas phase. Also, it is worth noting that the reaction rate coefficients of \ref{eq1} and \ref{eq2} are 1.9 $\times$ 10$^{-9}$ and 1.1 $\times$ 10$^{-9}$ cm$^{-3}$ s$^{-1}$, respectively \citep{Rakshit82,Adams1978}, which are within the typical values of fast protonation reactions of neutral species. Note that in this study, based on mass spectrometry, there is no differentiation between the two O- and S-protonated isomers. Alternatively, other protonation reactions, which exhibit slightly slower rate coefficients ($k$(\ref{eq3}) = 0.8 $\times$ 10$^{-10}$ cm$^{-3}$, $k$(\ref{eq4}) = 8.5 $\times$ 10$^{-10}$ cm$^{-3}$ and $k$(\ref{eq5})= 9.5 $\times$ 10$^{-10}$ cm$^{-3}$; \citealt{Smith78}), may take place as well:

\begin{equation}
 \mathrm{\ch{OCS}} + \mathrm{\ch{CH3+}}\rightarrow \rm \mathrm{\ch{HOCS+}} + \mathrm{\ch{CH3}} 
\label{eq3}
\end{equation}
\begin{equation}
 \mathrm{\ch{OCS}} + \mathrm{\ch{CH+}}\rightarrow \rm \mathrm{\ch{HOCS+}} + \mathrm{\ch{C}} 
\label{eq4}
\end{equation}
\begin{equation}
 \mathrm{\ch{SO}} + \mathrm{\ch{CH3+}}\rightarrow \rm \mathrm{\ch{HOCS+}} + \mathrm{\ch{H2}} 
\label{eq5}
\end{equation}

Once \ch{HOCS+} is formed, it can further recombine electronically to yield either CS and OCS \citep{Vidal17}. Furthermore, the large barrier to the isomerization of \ch{HSCO+} to \ch{HOCS+} (located at $\sim$70 kcal mol$^{-1}$ above \ch{HSCO+}, \citealt{Wheeler06}) will hamper any interconversion process. Hence, the observed isomeric ratio (\ch{HOCS+}/\ch{HSCO+}$\geq$2.3) can be rationalized in terms of the kinetics of the protonation pathways. A reasonable explanation was already put forward by \citet{McCarthy07}, who suggested that the arrangement of OCS with respect to the protonating agent, \ch{H3+}, shall play a crucial role in the protonation process. If this reaction is sensitive to the spatial orientation between the reactants, \ch{HOCS+} may prevail over \ch{HSCO+} in the ISM, because O is more electronegative than S and it will be the preferred binding site \citep{McCarthy07}. Likewise, in G+0.693 the observations suggest that O-protonation process is more efficient, which appears among the main chemical routes to the formation of \ch{HOCS+}.

In addition, most of the suggested precursors included in Figure \ref{f:model} (e.g., OCS, \ch{HCS+} and SO) are simple molecules that are present at relatively high abundances toward the G+0.693 molecular cloud (see e.g., SO with a column density of (30.06 $\pm$ 0.07) $\times$10$^{14}$ cm$^{-2}$ and an abundance of (2.2 $\pm$ 0.2) $\times$10$^{-8}$ with respect to H$_2$; \citealt{rivilla2022b}). We can thus compare the relative abundance of both protonated isomers (\ch{HOCS+} and \ch{HSCO+}) with respect to the parent carbonyl sulfide, OCS. This ratio appears as an excellent test for ion-molecule formation models since it relies significantly upon the fast protonation rates of the neutral molecules and upon fast dissociative electron recombination rates of the corresponding ion (i.e., \ch{HOCS+}; \citealt{Turner1990Sulfur}). We find a \ch{HOCS+}/OCS ratio of $\sim$2.5$\times$10$^{-3}$, while the \ch{HSCO+}/OCS upper limit ratio is $\leq$10$^{-3}$ (see Table \ref{tab:comparison}). Our results are in accordance with earlier observations toward warm star-forming regions, where upper limits of a few times 10$^{-3}$ were obtained for the \ch{HOCS+}/OCS ratio \citep{Turner1990Sulfur}. 

\begin{figure*}
\centerline{\resizebox{1.0\hsize}{!}{\includegraphics[angle=0]{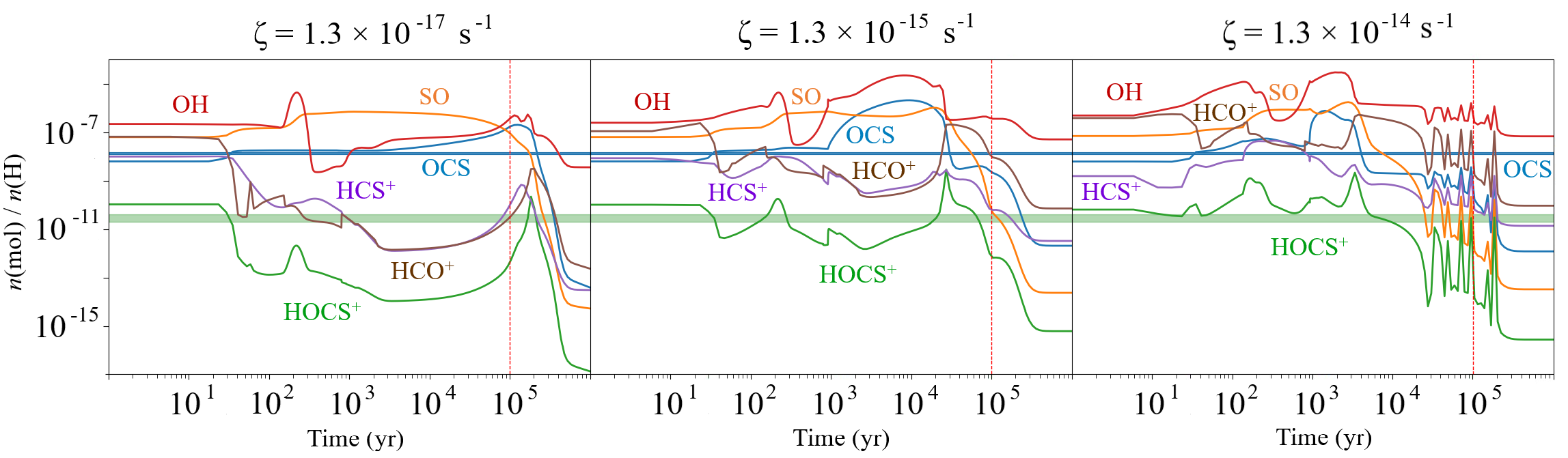}}}
\caption{Chemical model results (Phase 2): evolution of the fractional abundances of \ch{HOCS+} (green), OCS (blue) \ch{HCS+} (purple), \ch{HCO+} (brown), OH (red), and SO (orange) as a function of time. We consider for this phase a C-type shock with a shock speed of v$_s$=20 km s$^{-1}$ and an initial gas density of $n$(H) = 2$\times$10$^4$ cm$^{-3}$. We also explore three different values for the cosmic-ray ionisation rate: the standard Galactic value of $\zeta$ = 1.3$\times$10$^{-17}$ (left panel), and two enhanced values of $\zeta$$\times$100 (middle panel) and $\times$1000 (right panel) times higher than the standard one. The observed abundances of HOCS$^+$ and OCS, suggested as the main precursor, are highlighted using green and blue thick lines, respectively, adopting $n$(H)= 2 $\times$ $n$(H$_2$).}
\label{f:UCLCHEM}
\end{figure*}

We explored other feasible formation routes of \ch{HOCS+}, following what has been proposed for the formation of the isovalent \ch{HOCO+}, including the analogue protonation of desorbed \ch{CO2} and also the gas-phase route \ch{HCO+} + OH $\rightarrow$ \ch{HOCO+} + H (see e.g., \citealt{Fontani2018,Majumdar2018,Harada22}). Thus, if we assume a similar formation for the S-bearing species, we can propose the following ion-molecule reactions: 


\begin{equation}
\mathrm{\ch{HCS+}} + \mathrm{\ch{OH}} \rightarrow \rm \mathrm{\ch{HOCS+}} + \mathrm{\ch{H2}} 
\label{eq6}
\end{equation}
\begin{equation}
 \mathrm{\mathrm{\ch{HCO+}} + \ch{HS}}  \rightarrow \rm \mathrm{\ch{HOCS+}} + \mathrm{\ch{H2}} 
\label{eq7}
\end{equation}

Regarding the gas-phase route starting from \ch{HCS+} and OH (reaction \ref{eq6}), we find a molecular column density for \ch{HCS+} of (5.3 $\pm$ 2) $\times$ 10$^{13}$ (using entry 045506 of the CDMS catalog; \citealt{margules03}), which is $\sim$5 times more abundant than \ch{HOCS+}, and appears as another plausible precursor (the complete LTE analysis is shown in Appendix \ref{AnalysisHCS+}). Meanwhile, HS remains undetected toward G+0.693 due to the absence of spectroscopic features within the frequency ranged covered by our dataset. 

\subsection{Chemical modeling of \ch{HOCS+}}
\label{subsec:chemmod}

Initial models of the \ch{HOCS+} formation were carried out by \citet{Turner1990Sulfur} for typical molecular dark clouds conditions. For instance, we find a good agreement with the ``low-metal" gas-phase model estimates of \ch{HOCS+}/OCS gathered in Table 7 of \citet{Turner1990Sulfur} (e.g., for their model (3) the ratio is 3.3$\times$10$^{-3}$). However, these conditions are very different from the physical conditions of the molecular clouds located in the Galactic Center. Hence, to rationalize the observed abundances of HOCS$^+$ toward G+0.693, we have also employed the chemical code \textsc{UCLCHEM} \citep{holdship_uclchem:_2017}, as previously done for PO$^+$  \citep{rivilla2022b}. The model has been run in three different phases: Phase 0 simulates the chemistry of a translucent cloud exhibiting a $n$(H) = 10$^3$ cm$^{-3}$ and a $T_{kin}$ = 20 K for a period of 10$^6$ years. In Phase 1 we simulate the collapse of a molecular cloud from $n$(H) = 10$^3$ cm$^{-3}$ to $n$(H) = 2$\times$10$^4$ cm$^{-3}$ ($T_{kin}$ is set constant at 10 K). We then simulate in Phase 2 the passage of a low-velocity C-type shock with v$_s$=20 km s$^{-1}$ and pre-shock gas density of $n$(H) = 2$\times$10$^4$ cm$^{-3}$ using the C-type shock parametric approximation by \cite{jimenez-serra2008}. We adopt a shock velocity of v$_s$=20 km s$^{-1}$, consistent with the observed line widths of the molecular line emission \citep{requena-torres_organic_2006,zeng2018}, and with the gas densities measured toward G+0.693 \citep{zeng2020}. The initial elemental abundances are those used in \cite{Jiménez-Serra_2018}, adopting the solar value reported for S in \citet{Asplund:2009eu} (1.32$\times$10$^{-5}$; depletion factor of 1). This value falls within the range of (0.7-3.5)$\times$10$^{-5}$ reported for several GC sources \citep{Rodriguez-Fernandez2005}, as well as the S abundance found for other sources in the inner Galaxy of (0.7-1.6)$\times$10$^{-5}$ (e.g., (0.7$\pm$0.1)$\times$10$^{-5}$ for SgrC; \citealt{Martin-hernandez2002}). We also carried out models using larger depletion factors for S of about 4 (a total abundance of 3.51$\times$10$^{-6}$ from \citealt{Jenkins:2009ke}) and 10, respectively, which fail to reproduce the observed abundances of different S-bearing molecules, including HOCS$^+$ and OCS, even when considering an enhanced cosmic-ray ionization rate by several orders of magnitude (see below). 

Moreover, we employed the default dust-grain network of \textsc{UCLCHEM}, based on \cite{Quenard2018}, and the UMIST12 database \citep{mcelroy2013}. We have also introduced the gas phase reaction (\ref{eq6}), which to our knowledge has not been measured experimentally, assuming a rate constant of 10$^{-9}$ cm$^{-3}$, which is the typical value for ion-neutral reactions \citep{Puzzarini22}. 



To mimic the extreme physical conditions of G+0.693, which is thought to be affected by a strong ultraviolet field of secondary UV photons induced by cosmic rays, we have explored three distinct values for the cosmic-ray ionisation rate: the standard Galactic disk value of $\zeta$ = 1.3$\times$10$^{-17}$ s$^{-1}$, and two enhanced values of $\zeta$$\times$100 and $\times$1000 times higher than the canonical one. In Figure \ref{f:UCLCHEM} we show the model results for Phase 2, including the modelled fractional abundances of HOCS$^+$, OCS, HCS$^+$, HCO$^+$, SO and OH (solid coloured lines) along with the observed values for HOCS$^+$ and OCS, which is suggested as the main precursor (depicted with green and blue thick lines, respectively). The following results are derived from the models:

\begin{figure*}
\centerline{\resizebox{0.95\hsize}{!}{\includegraphics[angle=0]{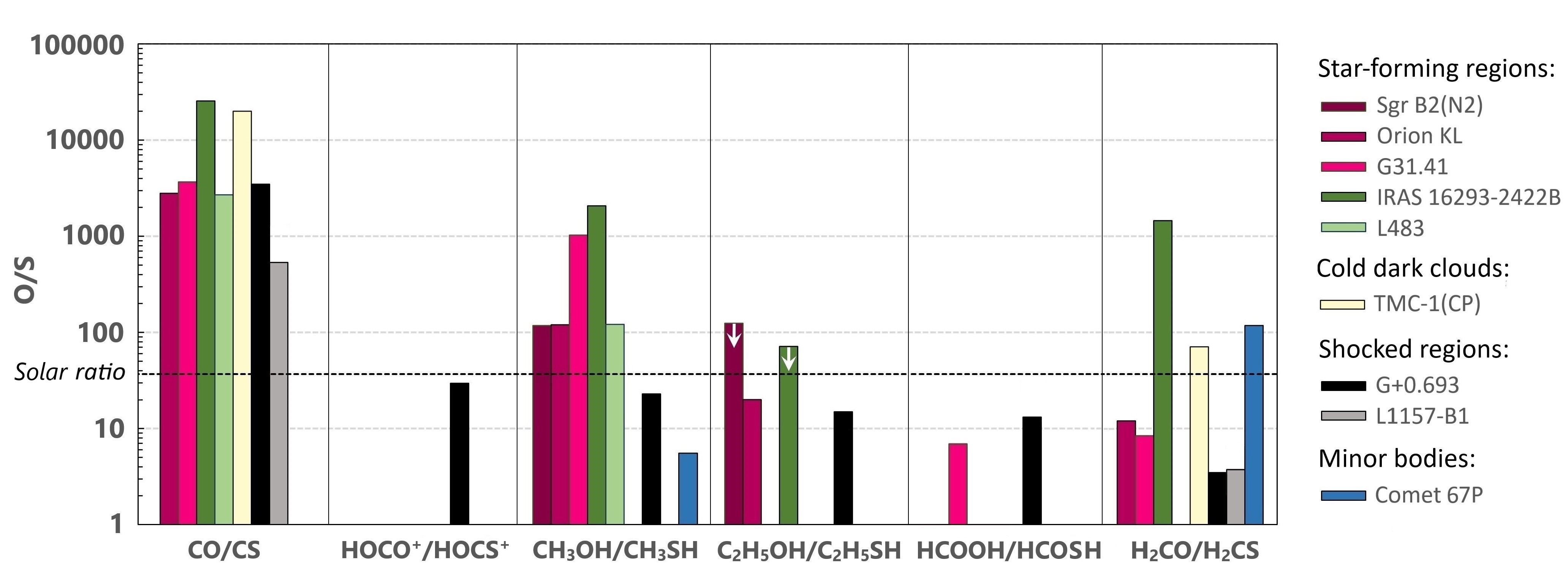}}}
\caption{Relative O/S ratio derived toward different interstellar sources for a sample of well-known S- and O-bearing species. Data are taken from: Sgr B2(N2) \citep{Muller:2016kd},  Orion KL \citep{Tercero:2010ft,Kolesnikova:2014fb} , G31.41 (López-Gallifa et al. submitted), IRAS 16293-2422B \citep{Drozdovskaya2019}, L483 \citep{agundez2019}, TMC-1 \citep{Walsh09}, L1157-B1 \citep{Holdship19}, G+0.693 (\citealt{zeng2018,rodriguez-almeida2021a} and this work) and comet 67P/C-G \citep{Drozdovskaya2019}. Upper limit ratios are shown with a white arrow.}
\label{f:sources}
\end{figure*}

\begin{enumerate}
\item We observe a release of OCS into the gas phase, which is ejected from the grains due to shocks, at time-scales of $\sim$40 yr once the sputtering of the icy mantles of dust grains is completed (see \citealt{jimenez-serra2008}).
\item It is clear that a high cosmic-ray ionisation rate (at least a factor of 100 higher than the standard value, left panel of Figure \ref{f:UCLCHEM}) is required to yield HOCS$^+$ abundances close to the observed value, in line with the results derived for PO$^+$ toward G+0.693 \citep{rivilla2022b}.
\item Although the gas phase reaction (\ref{eq6}) slightly increases the overall production of HOCS$^+$, the direct protonation of OCS stands as the dominant formation route of HOCS$^+$, specially for typical time-scales of Galactic center clouds (i.e., G+0.693) of about 10$^5$ yr \citep{RequenaTorres:2006ki}, marked by red dashed vertical lines in Figure \ref{f:UCLCHEM}. Note that to make the contribution of the reaction (\ref{eq6}) as relevant as the protonation process, we would need to increase its reaction rate by one order of magnitude ($k$ = 10$^{-8}$ cm$^{-3}$).
\item We also observed chemical relaxation oscillations for the highest $\zeta$ case (right panel of Figure \ref{f:UCLCHEM}). These oscillations are thought to occur due to the existence of bistable solutions \citep{Dufour19}, related to autocatalytic processes in the gas phase. Moreover, they can lead to limit cycles and chemical relaxation oscillation solutions when coupled with the gas-grain exchange of key species \citep{Dufour23}, and they usually take place at time-scales of interest for comparing models and observations.
\end{enumerate}


\subsection{Relative O/S ratio in different astronomical sources}
\label{subsec:soratio}

Understanding the S depletion problem has been an open subject of debate for astrochemical models \citep{Vidal17,Laas2019,shingledecker2020}. In low-density ($n$(H) $\leq$ 10$^{3}$-10$^{4}$ cm$^{-3}$; \citealt{Ruffle:1999fla}) and diffuse molecular clouds, the observed abundance of S with respect to that of O is close to the solar value (O/S$\sim$37; \citealt{Asplund:2009eu,Neufeld:2015boa,Goicoechea21}). On the contrary, in dense clouds ($n$(H) $\geq$ 10$^{5}$ cm$^{-3}$; \citealt{Tieftrunk94}) and protostars, the sulfur abundance found for the gas-phase chemical content is severely reduced up to several orders of magnitude, leading to the so-called ``missing sulfur" issue (e.g., \citealt{jimenez-escobar11,martin-domenech16,Marcelino23,Fuente23}). Thus, it is interesting to study the relative O/S ratio across different stages of star formation to establish general trends.

Hence, we present a compilation of the O/S ratio for a variety of well-known interstellar molecules, ranging from the CO/CS pair to the much more complex species like \ch{C2H5OH}/\ch{C2H5SH} toward various interstellar sources. We have included shock-dominated regions (L1157-B1, \citealt{Holdship19}, and G+0.693 \citealt{zeng2018,rodriguez-almeida2021a} and this work), a cold molecular cloud (TMC-1; \citealt{Walsh09}), high-mass (Sgr B2(N2), \citep{Muller:2016kd}, Orion KL, \citep{Tercero:2010ft,Kolesnikova:2014fb}, and G31.41 López-Gallifa et al. submitted) and low-mass (IRAS 16293-2442B, \citealt{Drozdovskaya2019}, and L483, \citealt{agundez2019}) star-forming regions, and in the comet 67P/Churyumov-Gerasimenko (hereafter 67P/C-G; \citealt{Drozdovskaya2019}). As shown in Figure \ref{f:sources}, among all different sources, G+0.693 appears as the one with the lowest O/S ratio, together with L1157-B1 and also comet 67P/C-G, which is thought to exhibit no S depletion \citep{Calmonte16}. Setting aside CO/CS, the O/S ratio derived toward G+0.693 follows a nearly steady factor of $\geq$10 (\ch{C2H5OH}/\ch{C2H5SH} = 15, \ch{CH3OH}/\ch{CH3SH} = 23 and also HCOOH/HCOSH = 13; \citealt{rodriguez-almeida2021a}). In this work, we obtained a \ch{HOCO+}/\ch{HOCS+} ratio of 31 $\pm$ 6, which agrees within a factor of 2-3 with the aforementioned pattern and also draws closer to the O/S solar ratio ($\sim$37; \citealt{Asplund:2009eu}). Only the CO/CS ratio is several orders of magnitude above the solar value, which may be due to the fact that CO traces a significative larger volume of gas than CS due to its lower critical density, or, alternatively, it suggests that the formation of CO is significantly more efficient than the production of CS toward G+0.693. These results can be rationalized in terms of the singular chemistry governing the cloud, which is subjected to large-scale shocks \citep{RequenaTorres:2006ki,zeng2020}. This translates into an enhanced sputtering erosion of the dust particles and a boost in the gas-phase abundance of diverse organics, including S-bearing molecules, rather than remaining anchored in the grains. Therefore, S appears to be relatively less depleted toward G+0.693 compared to other astronomical sources, further suggesting that the ``missing sulfur" problem is mitigated by the action of shocks, in agreement with the results presented in Sect. \ref{subsec:formation}.

Finally, based on the derived O/S ratios toward G+0.693, we can expect that the abundance of new S-bearing species will be at least an order of magnitude lower than the abundance of their O-bearing analogues (e.g., \ch{CH3OH} and \ch{C2H5OH}), in agreement with the lower limit ratio of $\geq$0.8-5.3 derived for the more complex \ch{NH2CH2CH2OH}/\ch{NH2CH2CH2SH} pair \citep{Song22} and also close to the solar ratio. Therefore, the recognition of O-bearing COMs with enough abundance will be an excellent guide for the selection of new S-bearing astronomical candidates of increasing complexity, fact that is intrinsically related to the cosmic availability of both elements.

\section{summary and conclusions} \label{sec:con}

We report the first detection of \ch{HOCS+}, the O-protonated form of carbonyl sulfide (OCS), in space based on an ultradeep spectral survey conducted toward the G+0.693-0.027 molecular cloud. We derived a molecular column density for \ch{HOCS+} of $N$ = (9 $\pm$ 2) $\times$ 10$^{12}$ cm$^{-2}$, which yields a fractional abundance with respect to H$_2$ of $\sim$7 $\times$ 10$^{-11}$. It is worth noticing that almost the entire set of transitions has been observed for the first time directly in the ISM and are still uncharted in the laboratory. Interestingly, we found that most of the lines of \ch{HOCS+} are partially blended with the emission from the $^{34}$S isotopologue of HNCS (HNC$^{34}$S), which, to the best of our knowledge, has been observed for the first time in the ISM in this work. Meanwhile, the S-protonated isomer, \ch{HSCO+} remains undetected, with an upper limit to its molecular abundance with respect to H$_2$ of $\leq$ 3 $\times$10$^{-11}$, at least 2.3 times less abundant than the O-protonated form. We have carried out new chemical models that account for the passage of a C-type shock, highlighting that high values of the cosmic-ray ionization rate ($\zeta$ = 10$^{-15}$-10$^{-14}$ s$^{-1}$) are required to explain the presence of \ch{HOCS+} in the G+0.693 molecular cloud. We also propose the protonation of OCS at the oxygen end to be one of the main chemical pathways to the formation of \ch{HOCS+} in the ISM.

We performed a comparison of the O/S ratio across different interstellar environments. In G+0.693, we obtained O/S values that are similar to the solar ratio (e.g., \ch{HOCO+}/\ch{HOCS+}$\sim$31, \ch{CH3OH}/\ch{CH3SH}$\sim$23 and \ch{C2H5OH}/\ch{C2H5SH}$\sim$15) which suggested that S is not significantly depleted. This is consistent with the idea that large-scale shocks in G+0.693 have released the ices of dust grains, boosting the abundance of diverse S-bearing species in the gas phase, instead of remaining trapped in the grains. Moreover, our results show that the exchange of the O atom by a S atom entails a drop of at least one order of magnitude in abundance. This fact can be used to guide the search for new S-bearing COMs of increasing complexity.

Finally, the detection of a new S-bearing cation provides valuable insights to disclose the role of ion-molecule processes in the formation of more complex interstellar systems, and will also aid theoretical chemists to update and verify both gas and gas-grain astrochemical networks targeting sulfur. Moreover, the discovery of \ch{HOCS+} reinforces the idea that the G+0.693 molecular cloud as a warehouse of not only N- and O- interstellar molecules but also of new S-bearing species.

\software{1) Madrid Data Cube Analysis (\textsc{Madcuba}) on ImageJ is a software developed at the Center of Astrobiology (CAB) in Madrid; \url{http://cab.inta-csic.es/madcuba/}; \citet{martin2019}; version from 2023 November 15.}

\begin{acknowledgments}
 
We are grateful to the IRAM 30$\,$m and Yebes 40$\,$m telescopes staff for their help during the different observing runs. The 40$\,$m radio telescope at Yebes Observatory is operated by the Spanish Geographic Institute (IGN, Ministerio de Transportes, Movilidad y Agenda Urbana). IRAM is supported by INSU/CNRS (France), MPG (Germany) and IGN (Spain). M.S.N. thanks the financial funding from the European Union - NextGenerationEU, Ministerio de Universidades and the University of Valladolid under a postdoctoral Margarita Salas Grant and also support from the Spanish Ministry of Science and Innovation (MCIN, PID2020-117742GB-I00). V.M.R. acknowledges support from project number RYC2020-029387-I funded by MCIN/AEI/10.13039/501100011033. I.J.-S., J.M.-P., L.C, A.M., and A.M.-H. acknowledge funding
from grants No. PID2019-105552RB-C41 and PID2022-136814NB-I00 from MCIN/AEI/10.13039/501100011033 and by “ERDF A way of making Europe”. A.M. has received support from project MDM-2017-0737-19-2, from grant PRE2019-091471, funded by MCIN/AEI/10.13039/501100011033 and by 'ERDF, A way of making Europe'. A.M.-H acknowledges funds from Grant MDM-2017-0737 Unidad de Excelencia “Mar{\'i}a de Maeztu" Centro de Astrobiolog{\'i}a (CAB, INTA-CSIC). V.M.R and D.S.A acknowledge the funds provided by the Consejo Superior de Investigaciones Cient{\'i}ficas (CSIC) and the Centro de Astrobiolog{\'i}a (CAB) through the project 20225AT015 (Proyectos intramurales especiales del CSIC). DSA also extends his gratitude for the financial support provided by the Comunidad de Madrid through the Grant PIPF-2022/TEC-25475. P.dV. and B.T. thank the support from MICIU through project PID2019-107115GB-C21. B.T. also thanks the Spanish MICIU for funding support from grant PID2022-137980NB-I00.

\end{acknowledgments}

\bibliography{rivilla,bibliography}{}
\bibliographystyle{aasjournal}

\newpage
\appendix
\twocolumngrid

\restartappendixnumbering

\section{Analysis of \ch{OCS} and its $^{34}$S isotopologue}
\label{AnalysisOCS}

\begin{figure*}
\centerline{\resizebox{1.0\hsize}{!}{\includegraphics[angle=0]{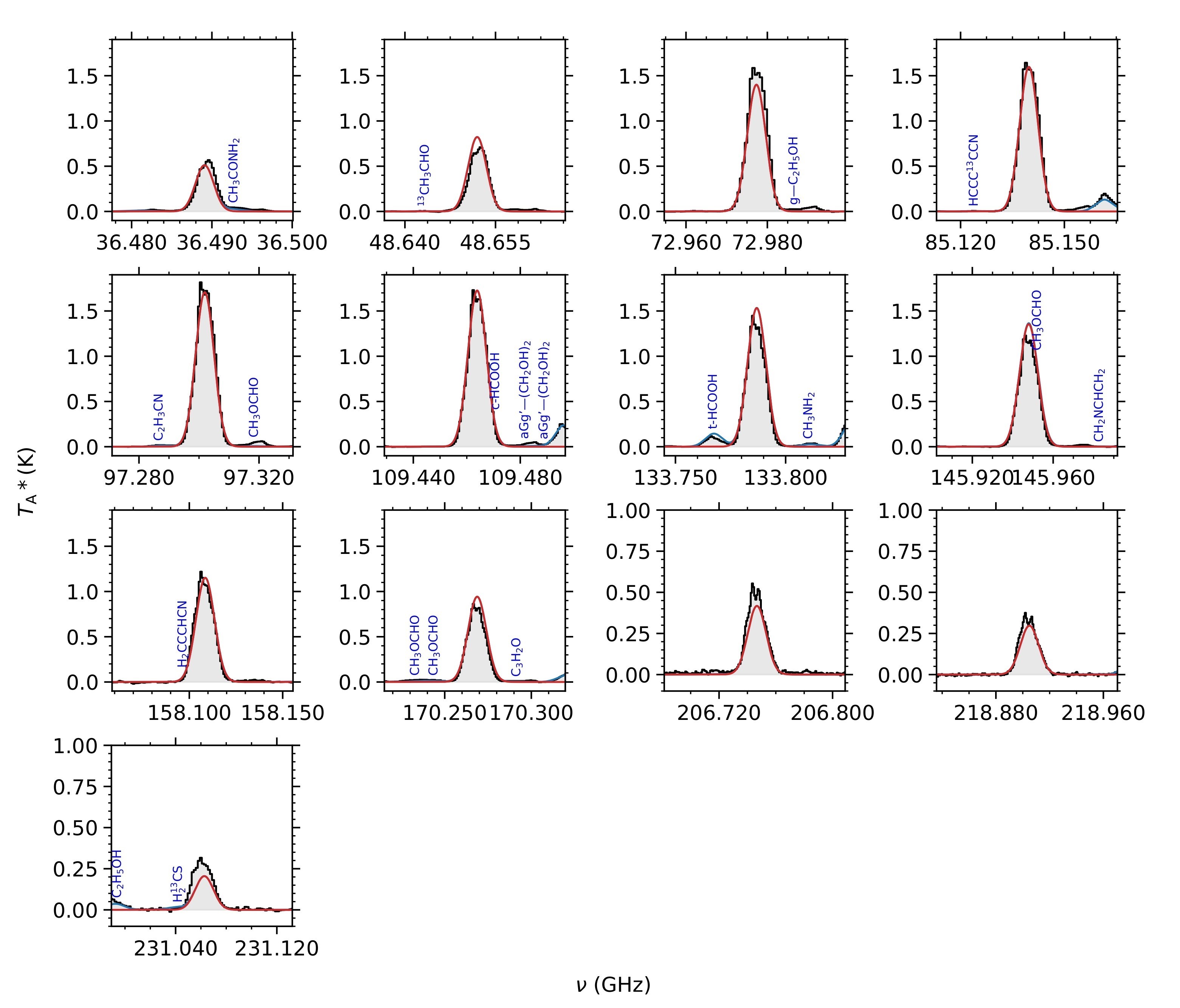}}}
\caption{Selected transitions of \ch{OCS} identified toward G+0.693 molecular cloud (listed in Table \ref{tab:ocs}). The result of the best LTE fit is shown with a red solid line, while the blue line shows the expected molecular emission from all the molecular species identified to date in our survey. The observed spectra are plotted as gray histograms.}
\label{f:OCS}
\end{figure*}

\begin{figure*}
\centerline{\resizebox{1.0\hsize}{!}{\includegraphics[angle=0]{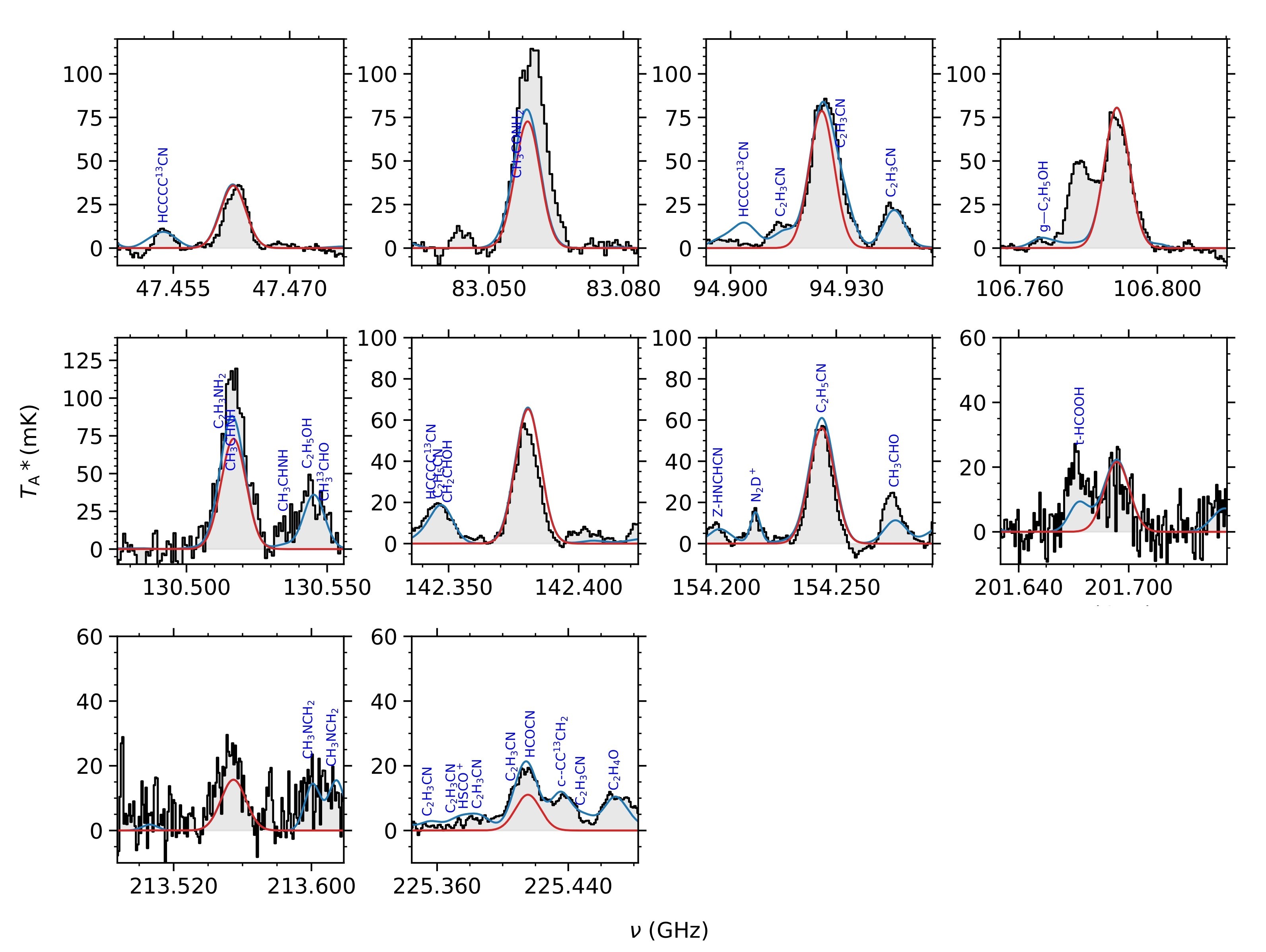}}}
\caption{Selected transitions of \ch{OC$^{34}$S} identified toward G+0.693 (listed in Table \ref{tab:ocs34}).}
\label{f:OCS34}
\end{figure*}

To conduct the LTE analysis of OCS we selected all the lines that are completely unblended with the emission from other species already identified toward G+0.693 (see Table \ref{tab:ocs}). In this case, since OCS is a linear molecule, we were able to identify harmonic patterns attributed to 13 unblended transitions. The result of the best LTE fit using \textsc{Autofit} tool within SLIM is shown in Figure \ref{f:OCS} and the derived physical parameter are reported in the first row of Table \ref{tab:comparison}. Then, we carried out the LTE analysis of its $^{34}$S monosubstituted isotopologue (see Figure \ref{f:OCS34} and Table \ref{tab:ocs34}). For OC$^{34}$S we obtained the following physical parameters in the LTE fit: $N$ = (1.80 $\pm$ 0.05) $\times$10$^{14}$ cm$^{-2}$, $T_{\rm ex}$ = (23.5 $\pm$ 0.6) K, $v$$_{\rm LSR}$ = (66.7 $\pm$ 0.3) km s$^{-1}$ and FWHM = (23.9 $\pm$ 0.7) km s$^{-1}$, in good agreement with the parameters derived for the parent OCS.  Thus, we derived a OC$^{32}$S/OC$^{34}$S ratio of 20 $\pm$ 1 toward G+0.693, which is very similar to the $^{32}$S/$^{34}$S obtained in the Galactic center by \citet{wilson_isotopes_1999} ($\sim$22). Hence, the emission of OCS toward G+0.693 appears to be optically thin, as further corroborated by the $^{16}$OCS/$^{18}$OCS ratio derived toward this cloud ($\sim$290, Colzi et al. in prep), which is in accordance with the $^{16}$OCS/$^{18}$OCS $\sim$250 found in \citet{armijos-abendano2015}.  The derived value is also similar to the \ch{CH3OH}/CH$_3$$^{18}$OH ratios of 210 $\pm$ 40 and  $\sim$180 reported in \citet{Gardner1989} and \citet{Muller:2016de} toward Sgr B2 and Sgr B2(N), respectively, as well as the $^{16}$O/$^{18}$O ratio presented in \citet{Wilson:1994wi} toward the Galactic center ISM.

\begin{table*}
\centering
\tabcolsep 3pt
\caption{Spectroscopic information of the selected transitions of \ch{OCS} detected toward G+0.693 (shown in Figure \ref{f:OCS}).}
\begin{tabular}{ccccccccccc}
\hline\hline
Frequency & Transition $^{(a)}$ & log \textit{I}& \textit{g}$\mathrm{_u}$ & $E$$\mathrm{_{up}}$ &  Blending \\ 
(GHz) & &  (nm$^2$ MHz) &  &  (K)  & \\
\hline
36.4888121  & 3-2  & -4.8824  & 7  & 3.5 &  Unblended \\
48.6516043  & 4-3  & -4.5105  & 9  & 5.8 &  Unblended \\
72.9767794  & 6-5  & -3.9907  & 13  & 12.2 &  Unblended \\
85.1391032  & 7-6  & -3.7954  & 15  & 16.2 &  Unblended \\
97.3012085  & 8-7  & -3.6277  & 17  & 20.9 &  Unblended \\
109.4630630  & 9-8  & -3.4815  & 19  & 26.1 &  Unblended \\
133.7859000  & 11-10  & -3.2369  & 23  & 38.3 &  Unblended \\
145.9468120  & 12-11  & -3.1333  & 25  & 45.2 &  Unblended \\
158.1073600  & 13-12  & -3.0396  & 27  & 52.7 &  Unblended \\
170.2674940  & 14-13  & -2.9544  & 29  & 60.9 &  Unblended \\
206.7451558  & 17-16  & -2.7408  & 35  & 88.7 &  Unblended \\
218.9033555   & 18-17  & -2.6811  & 37  & 99.1 &  Unblended \\
231.0609934  & 19-18  & -2.6263  & 39  & 110.1 &  Unblended \\
\hline 
\end{tabular}
\label{tab:ocs}
\vspace*{-2.5ex}
\tablecomments{$^{(a)}$ The rotational energy levels are labelled using the quantum number $J$.}
\end{table*}

\begin{table*}
\centering
\tabcolsep 3pt
\caption{Spectroscopic information of the selected transitions of \ch{OC$^{34}$S} detected toward G+0.693 (shown in Figure \ref{f:OCS34}).}
\begin{tabular}{ccccccccccc}
\hline\hline
Frequency & Transition $^{(a)}$ & log \textit{I}& \textit{g}$\mathrm{_u}$ & $E$$\mathrm{_{up}}$ &  Blending  \\ 
(GHz) & &  (nm$^2$ MHz) &  &  (K)  & \\
\hline
47.4623518  & 4 -- 3  & --4.5424  & 9  & 5.7 &  Unblended \\
83.0579710 & 7 -- 6  & --3.8269  & 15  & 15.8 &  Slightly blended: \ch{CH3CONH}  \\
94.9227994  & 8 -- 7  & --3.6591  & 17  & 20.4 &  Slightly blended: \ch{C2H3CN} \\
106.7873895  & 9 -- 8  & --3.5126 & 19  & 25.4 &  Unblended \\
130.5157300  & 11 -- 10  & --3.2677  & 23  & 37.3 &  Slightly blended: \ch{CH3CHNH} and \ch{C2H3NH2} \\
142.3794297  & 12 -- 11 & --3.1638  & 25  & 44.1 &  Unblended \\
154.2427800  & 13 -- 12  & --3.0698  & 27  & 51.5 &  Slightly blended: \ch{C2H5CN} \\
201.6919790  & 17 -- 16  & --2.7698  & 35  & 86.5 &  Slightly blended: $t$-HCOOH \\
213.5530610  & 18 -- 17  & --2.7098  & 37  & 96.7 &  Unblended \\
225.4136380  & 19 -- 18  & --2.6546  & 39  & 107.4 &  Blended: \ch{C2H3CN} and HCOCN  \\
\hline 
\end{tabular}
\label{tab:ocs34}
\vspace*{-2.5ex}
\tablecomments{$^{(a)}$ The rotational energy levels are labelled using the quantum number $J$.}
\end{table*}

\section{Analysis of \ch{HNCS}}
\label{AnalysisHNCS}

\begin{figure*}
\centerline{\resizebox{1.0\hsize}{!}{\includegraphics[angle=0]{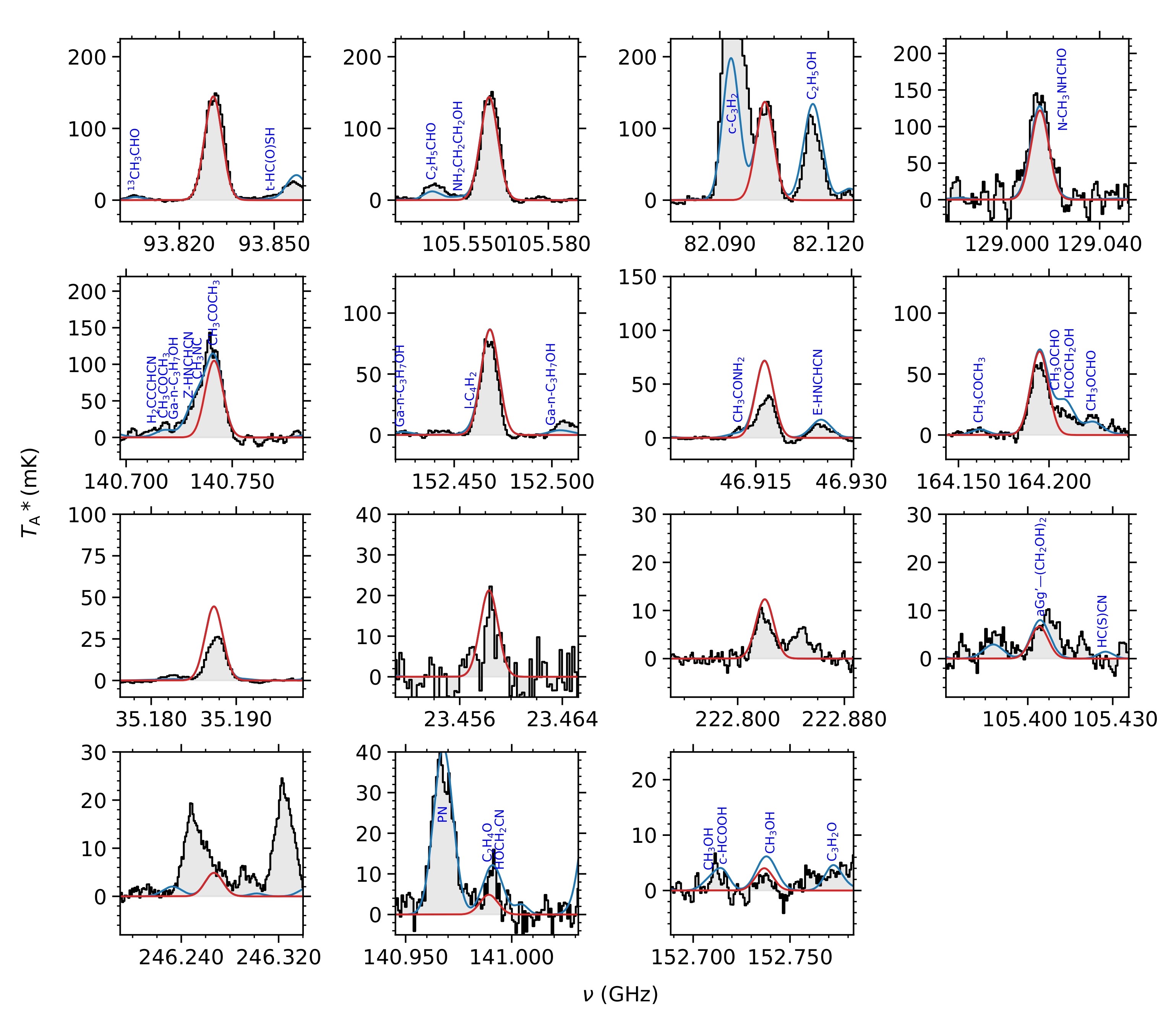}}}
\caption{Selected transitions of \ch{HNCS} identified toward G+0.693 (listed in Table \ref{tab:hncs}). The result of the best LTE fit is shown with a red solid line, the blue line shows the expected molecular emission from all the molecular species identified to date in our survey and the observed spectra are plotted as gray histograms.}
\label{f:LTEspectrumHNCS}
\end{figure*}

To carry out the analysis of HNCS under LTE conditions, we used entry 059503 of the CDMS catalog (corresponding to the $a$-type spectra). Then, we chose transitions that are unblended or slightly blended with the emission from other molecules to perform the fit using the SLIM \textsc{Autofit} tool \citep{martin2019}, which are collected in Table \ref{tab:hncs}. The result of the best LTE fit is shown in Figure \ref{f:LTEspectrumHNCS} and the derived physical parameter are: $N$ = (6.2 $\pm$ 0.1) $\times$10$^{13}$ cm$^{-2}$, which yields a fractional abundance with respect to molecular hydrogen of $\sim$4.6 $\times$ 10$^{-10}$ (adopting $N$(H$_{2}$) = 1.35$\times$10$^{23}$ cm$^{-2}$ from \citealt{martin_tracing_2008}), $T_{\rm ex}$ = (20.4 $\pm$ 0.5) K and $v$$_{\rm LSR}$ = (66.7 $\pm$ 0.3) km s$^{-1}$. The FWHM was fixed to 21.0 km s$^{-1}$ in the fit. These physical parameters are also collected in Table \ref{tab:comparison}.

\begin{table*}
\centering
\tabcolsep 3pt
\caption{Spectroscopic information of the selected transitions of \ch{HNCS} detected toward G+0.693 (shown in Figure \ref{f:LTEspectrumHNCS}).}
\begin{tabular}{ccccccccccc}
\hline\hline
Frequency & Transition $^{(a)}$ & log \textit{I}& \textit{g}$\mathrm{_u}$ & $E$$\mathrm{_{up}}$ &  Blending  \\ 
(GHz) & &  (nm$^2$ MHz) &  &  (K)  & \\
\hline
23.4581073  & 2$_{0,2}$ -- 1$_{0,1}$  & --5.3520  & 5  & 1.7 &  Unblended \\ 
35.1871100  & 3$_{0,3}$ -- 2$_{0,2}$  & --4.8258  & 7  & 3.4 &  Unblended \\
46.9160000  & 4$_{0,4}$ -- 3$_{0,3}$  & --4.4538  & 9  & 5.6 &  Slightly blended: \ch{CH3CONH2} \\
82.1018243  & 7$_{0,7}$ -- 6$_{0,6}$  & --3.7381  & 15  & 15.7 &  Slightly blended: $c$-\ch{C3H2} \\
93.8300500  & 8$_{0,8}$ -- 7$_{0,7}$  & --3.5703  & 17  & 20.1 &  Unblended \\
105.4032580 & 9$_{1,9}$ -- 8$_{1,8}$  & --3.5210  & 19  & 87.3 &   Slightly blended: \ch{(CH2OH)2}\\
105.5580740 & 9$_{0,9}$ -- 8$_{0,8}$  & --3.4238  & 19  & 25.2 &   Unblended \\
129.0132490 & 11$_{0,11}$ -- 10$_{0,10}$  & --3.1786  & 23  & 36.9 & Slightly blended: N-\ch{CH3NHCHO} \\
140.7403750 & 12$_{0,12}$ -- 11$_{0,11}$  & --3.0746  & 25 & 43.6 &  Slightly blended: \ch{CH3COCH3}, \ch{CH3NC} and $Z$-HNCHCN \\
140.9881310 & 12$_{1,11}$ -- 11$_{1,10}$  & --3.1668  & 25  & 105.8 &  Blended: \ch{C2H4O} and \ch{HOCH2CN} \\
152.4671300 & 13$_{0,13}$ -- 12$_{0,12}$  &  --2.9805 & 27  & 50.9 &  Slightly blended: l-\ch{C4H2} \\
152.7355410 & 13$_{1,12}$ -- 12$_{1,11}$  &  --3.0723 & 27  & 113.1 &  Slightly blended: \ch{CH3OH} \\
164.1935250 & 14$_{0,14}$ -- 14$_{0,14}$  & --2.8950  & 29  & 58.7 &  Blended with U \\
222.8186260 & 19$_{0,19}$ -- 18$_{0,18}$  & --2.5644  & 39  & 95.6 &  Unblended \\
246.2649890 & 21$_{0,21}$ -- 20$_{0,20}$  & --2.4666  & 43  & 129.1 &  Blended with U \\
\hline 
\end{tabular}
\label{tab:hncs}
\vspace*{-2.5ex}
\tablecomments{$^{(a)}$ The rotational energy levels are labelled using the conventional notation for asymmetric tops: $J_{K_{a},K_{c}}$, where $J$ denotes the angular momentum quantum number, and the $K_{a}$ and $K_{c}$ labels are projections of $J$ along the $a$ and $c$ principal axes.}
\end{table*}

\section{Analysis of \ch{HOCO+}}
\label{AnalysisHOCO}

\begin{figure*}
\centerline{\resizebox{0.66\hsize}{!}{\includegraphics[angle=0]{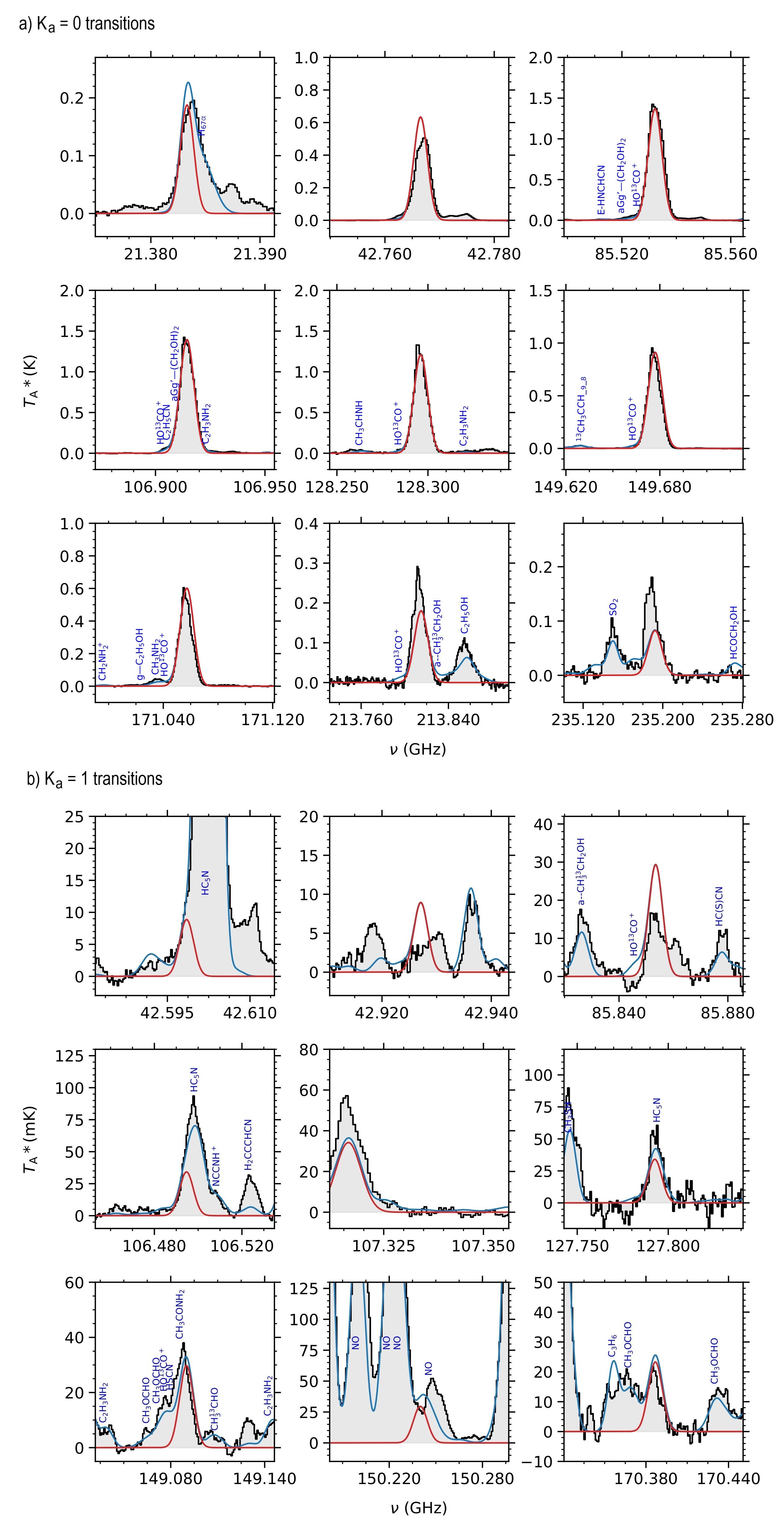}}}
\caption{Selected $K$$_a$ = 0 (a) and $K$$_a$ = 1 (b) transitions of \ch{HOCO+} identified toward G+0.693 (listed in Table \ref{tab:hoco}). The result of the best LTE fit is shown with a red solid line, while the blue line shows the expected molecular emission from all the molecular species identified to date in our survey. The observed spectra are plotted as gray histograms.}
\label{f:LTEspectrumHOCO}
\end{figure*}

To properly model the emission of \ch{HOCO+} under LTE conditions, we needed first to split the $K$$_a$ = 0, 1 rotational ladders. The selected transitions are listed in Table \ref{tab:hoco} while in Figure \ref{f:LTEspectrumHOCO}a ($K$$_a$ = 0) and \ref{f:LTEspectrumHOCO}b ($K$$_a$ = 1) we show the results of the best LTE fit using the \textsc{Autofit} tool within SLIM for each ladder. The derived physical parameters for the $K$$_a$ = 0 ladder are reported in Table \ref{tab:comparison}. As it can be seen, two $K$$_a$ = 1 transitions (panels 2 and 3 of Figure \ref{f:LTEspectrumHOCO}b) are still overestimated. This fact may be due to plausible non-LTE effects that are currently out of the scope of this work.

\begin{table*}
\centering
\tabcolsep 3pt
\caption{Spectroscopic information of the selected transitions of \ch{HOCO+} detected toward G+0.693 (shown in Figure \ref{f:LTEspectrumHOCO}).}
\begin{tabular}{ccccccccccc}
\hline\hline
Frequency & Transition $^{(a)}$ & log \textit{I}& \textit{g}$\mathrm{_u}$ & $E$$\mathrm{_{up}}$ &  Blending \\ 
(GHz) & &  (nm$^2$ MHz) &  &  (K)  & \\
\hline
21.3831524  & 1$_{0,1}$ -- 0$_{0,0}$  & --5.1233 & 3  & 1.0 &  Blended: H-67$\alpha$ \\ 
42.5981847  & 2$_{1,2}$ -- 1$_{1,1}$  & --4.4049 & 5  & 40.1 &  Blended: \ch{HC5N}  \\ 
42.7661888  & 2$_{0,2}$ -- 1$_{0,1}$  & --4.2224 & 5  & 3.1 & Unblended \\ 
42.9267850  & 2$_{1,1}$ -- 1$_{1,0}$  & --4.3982 & 5  & 40.2 &  Blended: U \\ 
85.5314968  & 4$_{0,4}$ -- 3$_{0,3}$  & --3.3283 & 9  & 10.2 &  Unblended  \\ 
85.8528356   & 4$_{1,3}$ -- 3$_{1,2}$  & --3.4072 & 9  & 47.3 &  Blended: U   \\ 
106.4938987  & 5$_{1,5}$ -- 4$_{1,4}$  & --3.1194 & 11  & 52.3 & Blended: HC(S)CN and \ch{HC5N}  \\ 
106.9135453  & 5$_{0,5}$ -- 4$_{0,4}$  &  --3.0442 & 11  & 15.3 & Blended: \ch{(CH2OH)2} and \ch{C2H5CN}  \\ 
107.3153561  & 5$_{1,4}$ -- 4$_{1,3}$  & --3.1128 & 11  & 52.4 & Unblended  \\ 
127.7916960  & 6$_{1,6}$ -- 5$_{1,5}$  & --2.8845 & 13  & 58.4 &  Slightly lended: \ch{HC5N} \\ 
128.2950630  & 6$_{0,6}$ -- 5$_{0,5}$  & --2.8149 & 13  & 21.4 &  Unblended \\ 
149.0889920   & 7$_{1,7}$ -- 6$_{1,6}$  & --2.6900 & 15  & 65.5 & Slightly blended: \ch{CH3CONH2}  \\ 
149.6758710  & 7$_{0,7}$ -- 6$_{0,6}$  & --2.6237 & 15  & 28.5 & Unblended  \\ 
150.2389283  & 7$_{1,6}$ -- 6$_{1,5}$  & --2.6836 & 15  & 65.7 &  Blended: NO \\ 
170.3855936  & 8$_{1,8}$ -- 7$_{1,7}$  & --2.5250 & 17  & 73.6 &  Unblended  \\ 
171.0559470  & 8$_{0,8}$ -- 7$_{0,7}$  & --2.4609  & 17  & 36.7 &   Unblended\\ 
213.8133390  & 10$_{0,10}$ -- 9$_{0,9}$  & --2.1969 & 21  & 60.0 &  Slightly blended: \ch{CH3$^{13}$CH2OH} \\ 
235.1904240  & 11$_{0,11}$ -- 10$_{0,10}$  & --2.0883 & 23  & 67.3 &  Blended: U \\ 
\hline 
\end{tabular}
\label{tab:hoco}
\vspace*{-2.5ex}
\tablecomments{$^{(a)}$ The rotational energy levels are labelled using the conventional notation for asymmetric tops: $J_{K_{a},K_{c}}$, where $J$ denotes the angular momentum quantum number, and the $K_{a}$ and $K_{c}$ labels are projections of $J$ along the $a$ and $c$ principal axes.}
\end{table*}

\section{Analysis of \ch{HCS+} and upper limit of the non-detected HCS}
\label{AnalysisHCS+}

\begin{figure*}
\centerline{\resizebox{0.7\hsize}{!}{\includegraphics[angle=0]{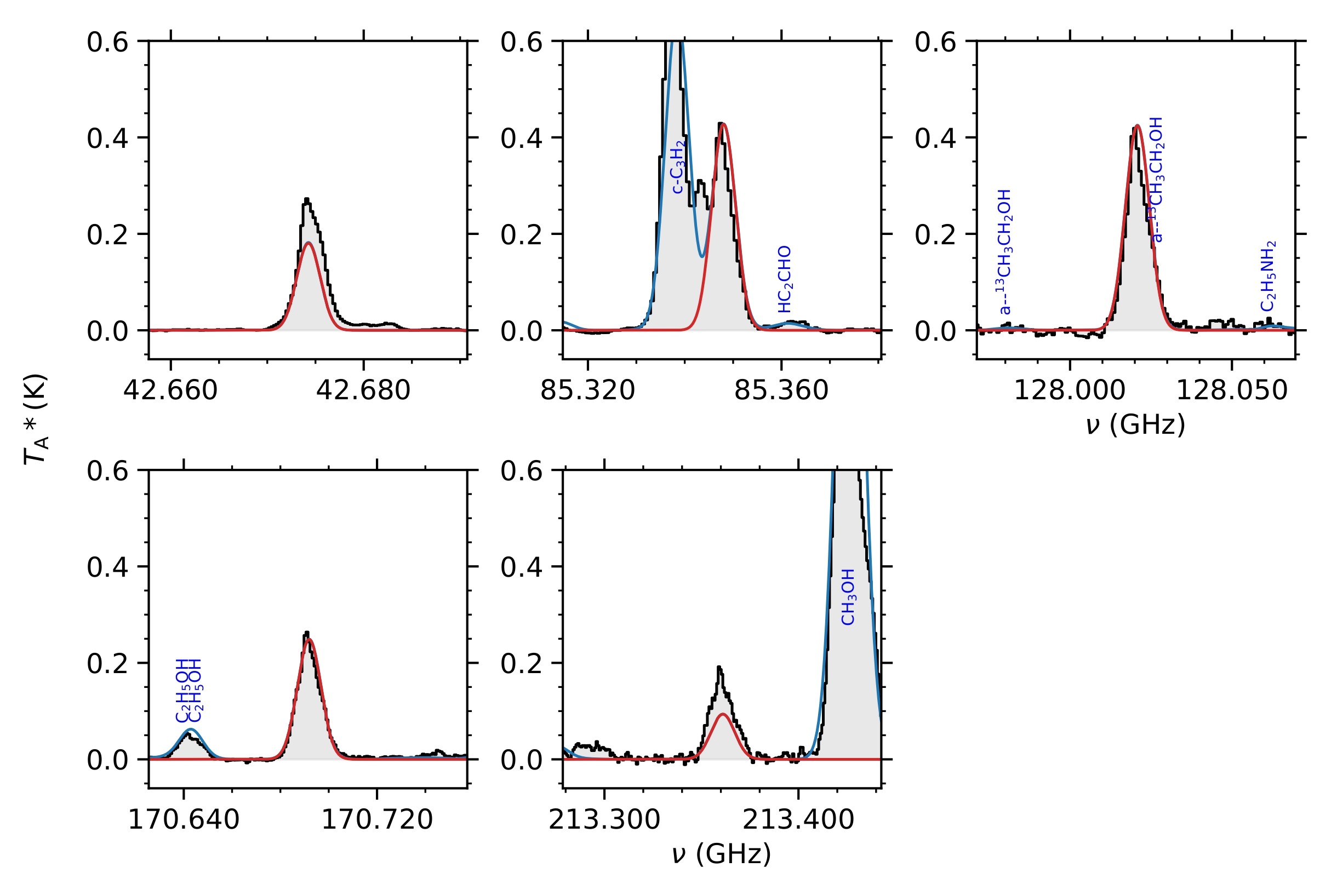}}}
\caption{Selected transitions of \ch{HCS+} identified toward the G+0.693 (listed in Table \ref{tab:hcs+}).}
\label{f:HCS+}
\end{figure*}

To perform the LTE analysis of \ch{HCS+}, a possible precursor of \ch{HOCS+}, we have employed entry 045506 of the CDMS catalog, whose spectroscopic information was taken from \citet{margules03}. We depict in Figure \ref{f:HCS+} the result of the best LTE fit using the \textsc{Autofit} tool within SLIM, including the five detected lines (a complete progression ranging from $J_{up}$ = 1 to $J_{up}$ = 5) listed in Table \ref{tab:hcs+}. We derived a molecular abundance of $N$ = (5.3 $\pm$ 2) $\times$ 10$^{13}$ cm$^{-2}$, which implies that \ch{HCS+} is $\sim$5 times more abundant than \ch{HOCS+}. Regarding the rest of physical parameters, these are: $T_{\rm ex}$ = (6.9 $\pm$ 0.2) K and $v$$_{\rm LSR}$ = (68.5 $\pm$ 0.2) km s$^{-1}$ and FWHM = (20.0 $\pm$ 0.6) km s$^{-1}$. Note that both the fundamental transition, $J$ = 1 -- 0, and the $J$ = 5 -- 4 are most likely unblended features affected by non-LTE effects, whose analysis is beyond the scope of this work.

\begin{table*}
\centering
\tabcolsep 3pt
\caption{Spectroscopic information of the transitions of \ch{HCS+} detected toward G+0.693 (shown in Figure \ref{f:HCS+}).}
\begin{tabular}{ccccccccccc}
\hline\hline
Frequency & Transition $^{(a)}$ & log \textit{I}& \textit{g}$\mathrm{_u}$ & $E$$\mathrm{_{up}}$ &  Blending  \\ 
(GHz) & &  (nm$^2$ MHz) &  &  (K)  & \\
\hline
42.6741954  & 1 -- 0  & --3.8017  & 3  & 2.0 &  Unblended \\
85.3478900 & 2 -- 1  & --2.9031  & 5  & 6.1 &  Slightly blended: $c$-\ch{C3H2}  \\
128.0205700  & 3 -- 2  & --2.3822  & 7  & 12.2 &  Slightly blended: $^{13}$\ch{CH3CH2OH} \\
170.6916030   & 4 -- 3  & --2.0178 & 9  & 20.3 &  Unblended \\
213.3606500  & 5 -- 4  & --1.7404  & 1  & 30.5 &  Unblended \\
\hline 
\end{tabular}
\label{tab:hcs+}
\vspace*{-2.5ex}
\tablecomments{$^{(a)}$ The rotational energy levels are labelled using the quantum number $J$.}
\end{table*}

\begin{figure}
\centerline{\resizebox{1.05\hsize}{!}{\includegraphics[angle=0]{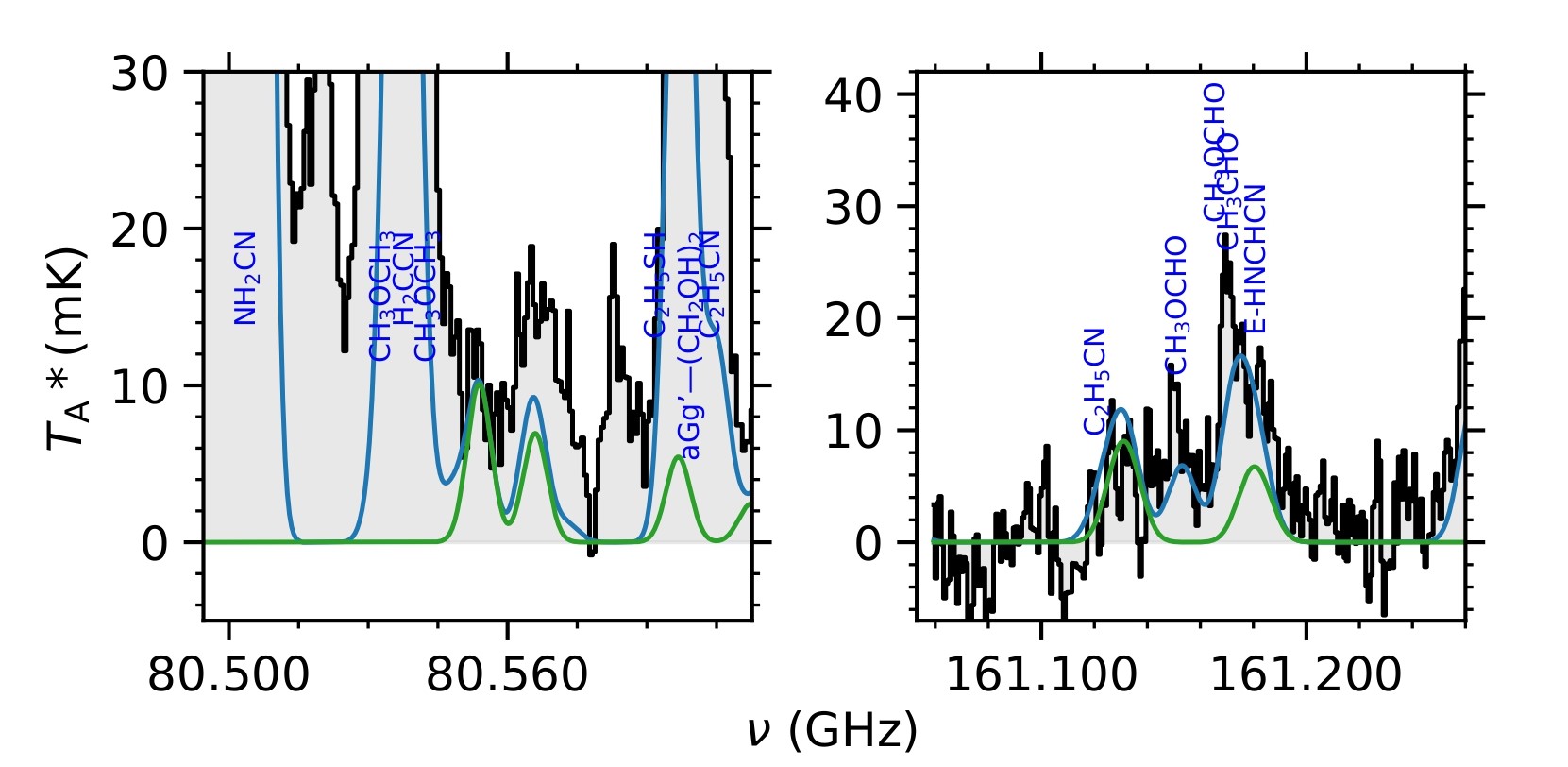}}}
\caption{LTE simulation of the HCS emission at the 3$\sigma$ upper limit column density derived toward G+0.693 (in green) together with the expected molecular emission from all the molecular species identified to date in our survey. The observed spectrum is shown as a gray histogram.}
\label{f:HCS}
\end{figure}

We have also searched for the HCS radical (CDMS entry 045507; \citealt{Habara:2002gd}), which is not clearly detected. Therefore, we have derived the 3$\sigma$ upper limit to its molecular abundance ($\sigma$ is the rms noise of the spectra) using the 2$_{0,2}$ -- 1$_{0,1}$ transition, in particular the $J$ = 5/2 -- 3/2 $F$ = 3 -- 2 hyperfine component, which is the brightest transition predicted in the LTE model (see Figure \ref{f:HCS}) that is fully unblended. We obtain a $N$ $\leq$ 9.8 $\times$ 10$^{13}$ cm$^{-2}$ using the physical parameters reported above for \ch{HCS+} (e.g., $T_{\rm ex}$ = 6.9 K). As seen in Figure \ref{f:HCS} some hyperfine components of the 2$_{0,2}$-1$_{0,1}$ and 4$_{0,4}$-3$_{0,3}$ transitions are tentatively detected, but there are not enough clear spectroscopic features to achieve a conclusive detection. We find that \ch{HCS} is $\leq$11 times more abundant than \ch{HOCS+} and $\leq$1.8 times more abundant than its cationic form, \ch{HCS+}.

\end{document}